\ifdefined\pdfminorversion\pdfminorversion=7\fi
\documentclass[11pt]{article}

\usepackage[margin=1in]{geometry}
\usepackage{amsmath,amssymb}
\usepackage{array}
\usepackage{booktabs}
\usepackage{float}
\usepackage{graphicx}
\usepackage{placeins}
\usepackage[colorlinks=true,linkcolor=blue,citecolor=blue,urlcolor=blue,
hyperfootnotes=false]{hyperref}
\usepackage[authoryear,round]{natbib}
\usepackage{setspace}
\usepackage{xcolor}

\graphicspath{{figures/}}

\newcommand{\flb}{favorite--longshot bias}

\title{The Favorite--Longshot Bias in Prediction Markets:\\
Evidence from Polymarket}

\author{Marcos Cardozo \qquad Jos\'e Ignacio Rivero-Wildemauwe\\[0.5em]
\small Universidad Cat\'olica del Uruguay\\
\small \texttt{marcos.cardozo@ucu.edu.uy} \qquad
\texttt{joseignacio.rivero@ucu.edu.uy}}

\date{September 2026}
\begin{document}

\maketitle

\begin{abstract}
\noindent The favorite--longshot bias (FLB) is one of the most persistent return
patterns in betting markets, but its prevalence in modern prediction markets
and its relation to recurrent longshot demand remain unclear. We study these
questions in Polymarket, a large and increasingly important prediction market,
using 588 million trades by 2.48 million accounts. In observed transaction flows,
purchases below 10 cents lose 19.3 cents per dollar, while purchases at or above
90 cents earn 0.83 cents. How contracts are grouped changes the result: 
longshots lose 6.3 cents per dollar when each contract is weighted equally but gain 4.1 
cents when related contracts are first grouped by parent event; the
two-sided pattern is robust in Crypto and Politics but surprisingly absent in Sports. The top decile of accounts with the strongest past tendency to buy longshots accounts for 26.6 percent of next-month low-price purchases, but earns similar returns to others. The top decile of accounts with the strongest past tendency to buy favorites accounts for 15.1 percent of next-month high-price purchases but earns less than others. These findings provide evidence of an aggregate FLB on Polymarket, but its magnitude depends on how contracts are grouped.

\end{abstract}

\section{Introduction}
\label{sec:introduction}
The favorite--longshot bias (FLB) is one of the oldest empirical regularities in betting markets: bets on unlikely outcomes tend to earn lower average returns than bets on more likely ones \citep{griffith1949,thalerziemba1988,snowbergwolfers2010}. Yet an aggregate FLB can conceal substantial differences in longshot and favorite returns and in who places those bets. Is the pattern common across different types of bets, or does it disappear or reverse in some? How much does the overall result depend on markets that account for the most spending or events that offer many related bets? Do longshots and favorites attract a recurring clientele, and do those bettors bear a disproportionate share of the associated losses and gains relative to their spending? 

In this paper we document the FLB at both price tails and examine these questions. Explaining the FLB requires more than reproducing the aggregate return pattern. An explanation must also be consistent with where the bias appears and how the associated gains and losses are distributed among bettors. Our findings allow us to test specific implications of existing explanations and can narrow the range of accounts consistent with the empirical evidence.

We use data from Polymarket, a large online prediction market. Participants trade YES--NO tokens on verifiable future events, such as the winner of an election. A token pays one dollar if its stated outcome occurs and zero otherwise, and it can be traded before the outcome is known \citep{polymarketdocs2026ctf}. The platform groups related questions under common headings that it calls ``events'' \citep{polymarketdocs2026marketsevents}. We call each separately traded question a \emph{child market} and the event under which it is listed its \emph{parent event}. For example, the parent event ``Presidential Election Winner 2024'' contained separate child markets asking whether Donald Trump, Kamala Harris, or another candidate would win. The transaction records also allow us to follow trading accounts, which we call \textit{wallets}, over time.\footnote{A person may use more than one wallet, so distinct wallets do not necessarily represent distinct individuals.}

We use the Polymarket Users dataset made available by \citet{akey2026}, specifically version 1.3 \citep{gregoire2026data}. It covers approximately 588 million trades by 2.48 million wallets from November 2022 through March 2026. For purchases in resolved markets, we compare the price paid with the amount the token ultimately pays. We use two primary calculations, applied separately to longshots and favorites. First, we calculate a return in each child market and give every child market equal weight. Second, we pool all purchases, so the calculation describes the return on the money invested. As a secondary aggregation approach, we pool purchases within each parent event and then give every parent event equal weight.

When each child market receives equal weight, purchases below 10 cents lose 6.30 percent, while purchases at or above 90 cents earn 0.277 percent; both estimates are statistically significant at the 0.1 percent level. When purchases are weighted by the dollars paid, longshots lose 19.35 percent and favorites earn 0.83 percent. The pooled longshot estimate is economically large, although its 95 percent confidence interval includes zero. In contrast, when each parent event receives equal weight, longshots earn 4.09 percent, while favorites continue to earn positive returns. Thus, the same purchases produce the classic FLB across child markets and invested dollars but not across parent events.

The reversal reveals two distinct forms of concentration. Parent events containing many child markets tend to have lower longshot returns and therefore account for the negative equal-child-market estimate. Separately, the parent events attracting the most longshot spending tend to have lower returns and account for the negative pooled estimate. The decomposition therefore shows how both the number of child markets and the concentration of spending shape aggregate longshot returns.

The pattern also differs by category. Crypto and Politics display longshot losses and favorite gains under both primary calculations. Sports is the clearest exception: longshots earn positive returns under both calculations, while favorite returns are not consistently positive. Results for Finance, Culture, Tech, and Weather depend on how purchases are weighted, and the longshot return in Weather changes sign.

At the wallet level, we use only previous purchases to identify wallets with unusually strong longshot or favorite demand. We then ask whether these buying patterns persist, how the wallets’ subsequent returns compare with those of other wallets, and whether their shares of gains and losses are large relative to their shares of spending.

Past purchases strongly predict subsequent purchases at both price tails. However, recurrent buyers do not account for disproportionate shares of gains or losses relative to their spending. Recurrent longshot buyers earn approximately the same negative return as other longshot buyers, while recurrent favorite buyers earn less than other favorite buyers. Repeated purchases at the price tails therefore do not identify groups that bear disproportionate shares of longshot losses or receive disproportionate shares of favorite gains.

Finally, we test whether longshot losses are confined to trading conditions that could help sustain them. We find that hey appear at every experience level and under all purchase methods, including when buyers post their own offers. They also persist in markets without reported fees and in events where collateral can be shared across mutually exclusive outcomes, reducing the cash needed to sell those tokens. The losses are therefore not confined to inexperienced wallets, purchases made by accepting sellers' offers, or markets with reported fees. Nor are they smaller in events where collateral can be shared, as a simple explanation based on collateral costs would suggest.

Existing explanations of the FLB emphasize mistaken probability judgments, preferences for risky payoffs, differences in beliefs or information, and features of market design \citep{weitzman1965,ali1977,ottavianisorensen2008,snowbergwolfers2010}. Prior work also shows that the strength of the bias varies with time to resolution and bettor experience \citep{pageclemen2013,feessmullerschumacher2014}. Our evidence does not identify a single explanation, but it narrows the possibilities: longshot losses persist across experience levels and purchase methods, without reported fees, and where sellers need less cash to take the opposite side. Moreover, recurrent longshot buyers do not bear losses out of proportion to their spending, weighing against a simple account based on a stable group of especially loss-prone longshot bettors.

Recent work on modern prediction markets shows that aggregate price accuracy can coexist with substantial heterogeneity across traders and trades. On Polymarket, \citet{reichenbach2025} study calibration over the market lifecycle; \citet{gomezcram2026} identify a small group of persistently skilled traders whose trading is associated with more accurate subsequent prices; and \citet{akey2026} document concentrated wallet profits and an advantage to liquidity provision. On Kalshi—another prediction market—\citet{becker2026} decomposes wealth transfers by order-book role and contract direction, while \citet{burgidengwhelan2026} document a favorite--longshot bias and model its variation across makers and takers. We complement this work by showing that the measured longshot return depends on the unit of aggregation. The same purchases yield negative longshot returns when each child market receives equal weight and when dollars are pooled, but positive returns when related contracts are grouped into broader events and those events receive equal weight. We then separate persistent demand from the incidence of these returns: wallets that repeatedly buy at the price tails do not receive gains or bear losses out of proportion to their spending. Thus, an aggregate favorite--longshot bias need not be broad across events, and recurrent longshot buying need not identify the wallets bearing disproportionate losses.

The remainder of the paper proceeds as follows. Section \ref{sec:data} describes the data and return measures. Section \ref{sec:exists} documents returns and price accuracy at the two price tails and explains the reversal across aggregation methods. Sections \ref{sec:generality} and \ref{sec:explanations} examine variation across categories and trading conditions, and Section \ref{sec:traders} studies recurrent longshot and favorite buyers. Section \ref{sec:interpretation} relates the findings to leading explanations of the FLB, and Section \ref{sec:conclusion} concludes.
\section{Data and measurement}
\label{sec:data}

\subsection{Polymarket and the dataset}

On Polymarket, participants trade with one another rather than against the
platform. A market is \emph{resolved} once its outcome has been determined
and the final token payoffs are known \citep{polymarketdocs2026ctf}.
Parent events differ in how they group child markets: some contain
alternative answers to one question, while others contain related questions
for which more than one answer can be correct
\citep{polymarketdocs2026marketsevents,akey2026,gregoire2026data}.

Participants trade through accounts that Polymarket calls \emph{wallets}. A
wallet is the unit we observe, but it need not correspond to one person, as one
person may use several wallets, and several people may share the same wallet. A
participant trades by submitting an order that states the token, quantity,
price, and whether she wants to buy or sell. The participant whose order is
waiting is the \emph{maker}; the participant who accepts it is the
\emph{taker}. The buyer receives the tokens and the seller receives payment
\citep{polymarketdocs2026overview}.\footnote{During our sample period, payments were made in
	USD Coin (USDC), which is designed to maintain a value of one U.S. dollar \citep{akey2026}. We
	therefore report prices and amounts using a dollar sign.}

We use version 1.3 of the Polymarket Users dataset
\citep{gregoire2026data}. The dataset reconstructs completed trades from
Polymarket's public records and links them to the corresponding market
descriptions and outcomes. It covers November 11, 2022 through March 29, 2026
and contains 588 million trades by 2.48 million wallets. For each trade, we
observe when it occurred, the market and token traded, the price and number of
tokens, the buyer's and seller's wallets, and which participant was the maker.
For each market, we observe its wording and category, its parent event, its opening
and closing dates, and the winning token if the market had resolved by the end
of the sample.

We use the market-category labels supplied with version 1.3 of the dataset. The seven named categories are ``Sports,'' ``Crypto,'' ``Finance,'' ``Politics,'' ``Tech,'' ``Culture,'' and ``Weather.'' The dataset also includes an ``Untagged'' label \citep{gregoire2026data}.

For each wallet, we also observe an estimate of its total gain or loss across markets resolved by March 29, 2026 \citep{gregoire2026data}.\footnote{The estimate adds the amounts received from token sales and payouts on winning tokens, then subtracts the amounts paid for token purchases. Unless otherwise stated, this is the measure we use when discussing gains and losses at the wallet level.} In Appendix \ref{app:incidence}, we use this measure to examine whether wallets classified as recurrent longshot buyers, taken together, lose money across their trading in resolved markets.

\subsection{Construction and coverage of the analysis sample}

The complete dataset contains 729,133 child markets in 316,429 parent events
and 2,480,104 distinct wallets. Of these listings, 614,883 child markets in
255,427 parent events record at least one trade during the sample.

We retain completed trades for which we can identify the maker, taker, and buyer and observe the price and number of tokens. We then exclude self-trades, zero or negative quantities, prices at or below zero or above one dollar, and markets with more than two outcome tokens.\footnote{Our
	tabulation of the released market metadata identifies 39 such markets, or
	approximately 0.005 percent of the 729,133 markets in the complete dataset.
	We exclude them from all analyses \citep{gregoire2026data}.} To reduce the influence of potential wash trading, we also exclude purchases
made by buyers whose counterparty Herfindahl--Hirschman index is at least
0.5.\footnote{Wash trading consists of trades intended to create the
	appearance of market activity without meaningfully changing who carries the
	financial risk. A high counterparty index means that a large share of a
	wallet's trading volume is concentrated among a small number of other wallets.
	It is therefore an
	indicator of potential wash trading. We apply the
	restriction to the \textit{buyer}. Therefore, purchase is not excluded solely because the seller
	crosses the threshold. Unlike \citet{akey2026}, which applies the threshold
	only to wallets with at least 100 trades in resolved markets, we apply it to
	all buyers.}

Each completed trade appears once in our analysis, as a purchase by the buyer. If, say, wallet A buys tokens and later sells them to wallet B, we record a purchase by A and another by B. Our purchase counts and dollar totals therefore measure how much trading occurs, not how many positions are created.

Table~\ref{tab:sample-construction} summarizes the construction of the two
samples used in the paper. Panel A follows the cumulative restrictions used to
construct the analysis sample and then the return sample. Panel B describes
the counterparty-HHI restriction at the wallet level.

\begin{table}[H]
	\centering
	\caption{Sample construction}
	\label{tab:sample-construction}
	\small
	\resizebox{\textwidth}{!}{%
		\begin{tabular}{@{}lrrrrrr@{}}
			\toprule
			\multicolumn{7}{@{}l}{\textit{Panel A. Construction of the purchase sample}} \\
			\addlinespace
			Sample
			& Purchases
			& \shortstack{Share of all\\transactions}
			& \shortstack{Child\\markets}
			& \shortstack{Parent\\events}
			& \shortstack{Amount\\paid}
			& \shortstack{Amount\\retained} \\
			\midrule
			Observed transactions
			& 588,287,492 & 100.0\% & 614,883 & 255,427 & $\$24.63$ billion & 100.0\% \\
			Complete fields; distinct buyer and seller
			& 588,286,587 & 100.0\% & 614,863 & 255,425 & $\$24.63$ billion & 100.0\% \\
			Valid quantity and price
			& 588,286,587 & 100.0\% & 614,863 & 255,425 & $\$24.63$ billion & 100.0\% \\
			Two-outcome markets
			& 588,258,033 & 100.0\% & 614,863 & 255,425 & $\$24.62$ billion & 100.0\% \\
			Buyer counterparty HHI $<0.5$
			& 586,123,163 & 99.6\% & 614,848 & 255,418 & $\$23.67$ billion & 96.1\% \\
			Resolved by March 29, 2026
			& 560,909,624 & 95.3\% & 591,187 & 250,307 & $\$22.49$ billion & 95.0\% \\
			\bottomrule
		\end{tabular}%
	}
	
	\medskip
	\begin{tabular*}{\textwidth}{@{\extracolsep{\fill}}lrrrr@{}}
		\toprule
		\multicolumn{5}{@{}l}{\textit{Panel B. Buyer counterparty concentration}} \\
		\addlinespace
		Wallet group
		& Wallets
		& \shortstack{Share of\\wallets}
		& \shortstack{Amount paid\\as buyer}
		& \shortstack{Share of\\amount paid} \\
		\midrule
		All wallets
		& 2,480,104 & 100.0\% & $\$24.62$ billion & 100.0\% \\
		Counterparty HHI $\geq0.5$
		& 439,876 & 17.7\% & $\$0.96$ billion & 3.9\% \\
		Counterparty HHI $<0.5$
		& 2,040,228 & 82.3\% & $\$23.67$ billion & 96.1\% \\
		\bottomrule
	\end{tabular*}
	
	\medskip
	\begin{minipage}{0.96\linewidth}
		\footnotesize
		Notes: Restrictions are cumulative. Each completed trade is counted once, as
		a purchase by the buyer. In Panel A, shares are relative to all observed
		transactions. Amount paid is calculated as $\sum_f p_fq_f$ at every stage. The
		amount retained is the share of the amount paid in the immediately preceding
		row. The counterparty-HHI restriction is applied to the buyer: a purchase is excluded when
		the buyer's counterparty HHI is at least 0.5. It is not excluded solely because
		the seller crosses that threshold. The final row of Panel A contains the
		purchases used to calculate returns. In Panel B, wallet counts refer to all
		wallets in the released data. Amounts are for purchases made by those wallets
		as buyers in two-outcome child markets before and after applying the
		counterparty-HHI restriction; the
		excluded amount is the difference between them.
	\end{minipage}
\end{table}

These restrictions leave an analysis sample of 586.1 million purchases
totaling \$23.67 billion. The return sample consists of purchases in markets
resolved by March 29, 2026, accounting for 95.7 percent of analysis-sample
purchases and 95.0 percent of the amount paid. Purchases in unresolved markets
remain in our descriptions of trading activity but do not enter the return
calculations.

Wallets above the counterparty-HHI threshold represent 17.7 percent of
wallets in the released data, but their purchases account for only
3.9 percent of the amount paid after the transaction and market restrictions.

It should be noticed that coverage is lower near the end of the sample because recent markets had less
time to resolve. For example, a purchase made in January 2024 had more than two
years to reach a final payoff by March 29, 2026, whereas a purchase made in
March 2026 had at most a few weeks. If the latter market was still unresolved
at the cutoff, its purchases do not enter the return sample. Return estimates
for recent purchases therefore describe purchases in markets that resolved
quickly, not necessarily all purchases made during the same period. We report the share of
the amount paid with a known payoff by purchase quarter alongside the
time-specific estimates. Table~\ref{tab:category-composition} shows how purchases, amounts paid, and
coverage of final payoffs vary across categories.

\begin{table}[H]
	\centering
	\caption{Analysis sample by category}
	\label{tab:category-composition}
	\small
	\resizebox{\textwidth}{!}{%
		\begin{tabular}{@{}lrrrrr@{}}
			\toprule
			category
			& Purchases
			& Share of purchases
			& Amount paid
			& Share of amount paid
			& \shortstack{Share of amount\\with known payoff} \\
			\midrule
			All categories      & 586.1M & 100.0\% & $\$23.67$B & 100.0\% & 95.0\% \\
			\midrule
			Politics        & 54.8M  & 9.3\%   & $\$6.74$B  & 28.5\%  & 91.0\% \\
			Sports          & 94.6M  & 16.1\%  & $\$8.72$B  & 36.8\%  & 97.0\% \\
			Crypto  & 394.1M & 67.2\%  & $\$5.97$B  & 25.2\%  & 98.2\% \\
			Finance         & 9.1M   & 1.6\%   & $\$806.1$M & 3.4\%   & 87.7\% \\
			Culture         & 19.4M  & 3.3\%   & $\$1.01$B  & 4.3\%   & 92.6\% \\
			Tech      & 5.8M   & 1.0\%   & $\$260.2$M & 1.1\%   & 90.5\% \\
			Weather         & 8.3M   & 1.4\%   & $\$155.5$M & 0.7\%   & 98.2\% \\
			\bottomrule
		\end{tabular}%
	}
	
	\medskip
	\begin{minipage}{0.96\linewidth}
		\footnotesize
		Notes: The table describes the analysis sample after the transaction, market,
		and buyer restrictions in Table~\ref{tab:sample-construction}. It includes
		purchases in both resolved and unresolved markets. Amount paid is
		$\sum_f p_fq_f$. The share with a known payoff is the amount paid in markets
		resolved by March 29, 2026 divided by the total amount paid in the
		corresponding category.
	\end{minipage}
\end{table}

The number of purchases and the amount paid are distributed differently across
categories. Crypto accounts for 67.2 percent of purchases but only 25.2 percent
of the amount paid. Sports and Politics together account for just 25.4 percent
of purchases but 65.3 percent of the amount paid. The share of the amount paid
for which a final payoff is known also varies across categories, from 87.7 percent
in Finance to 98.2 percent in Crypto and Weather. These differences
justify reporting the return estimates separately by category.

The 614,883 traded child markets belong to 255,427 parent events, but parent
events differ in the number of child markets they contain.
Table~\ref{tab:event-menu-distribution} reports the minimum, mean, median,
and maximum number of child markets per event. This distinction matters because averaging child
markets separately gives multi-market parent events repeated weight, whereas
averaging parent events gives each represented event one observation.

\begin{table}[H]
	\centering
	\caption{Distribution of child markets per parent event}
	\label{tab:event-menu-distribution}
	\small
	\begin{tabular*}{\textwidth}{@{\extracolsep{\fill}}lrrrrrr@{}}
		\toprule
		category
		& Child markets
		& Parent events
		& Minimum
		& Mean
		& Median
		& Maximum \\
		\midrule
		All categories & 614,883 & 255,427 & 1 & 2.41 & 1 & 144 \\
		Politics   & 29,926  & 7,248   & 1 & 4.13 & 1 & 70  \\
		Sports     & 321,231 & 64,205  & 1 & 5.00 & 3 & 144 \\
		Crypto     & 206,491 & 172,313 & 1 & 1.20 & 1 & 100 \\
		Finance    & 17,159  & 5,539   & 1 & 3.10 & 1 & 33  \\
		Culture    & 15,806  & 2,530   & 1 & 6.25 & 5 & 66  \\
		Tech       & 4,308   & 1,016   & 1 & 4.24 & 1 & 47  \\
		Weather    & 19,962  & 2,576   & 1 & 7.75 & 7 & 11  \\
		\bottomrule
	\end{tabular*}
	
	\medskip
	\begin{minipage}{0.96\linewidth}
		\footnotesize
		Notes: The last four columns summarize, within each category, the distribution
		of the number of distinct traded child markets per parent event. Means and
		medians give each parent event equal weight.
	\end{minipage}
\end{table}

\FloatBarrier

\subsection{Measuring returns}
\label{sec:estimands}

The return on a purchase depends on the price paid and the token's final
payoff. Consider again a participant who bought one YES token for 8 cents on
Donald Trump winning the 2024 presidential election. Because Donald Trump won,
the token paid one dollar. The buyer gained 92 cents on the 8 cents paid, a
return of 1,150 percent. Had Donald Trump lost, the token would have paid nothing
and the buyer would have lost the 8 cents paid, a return of $-100$ percent.

We apply this calculation to every purchase in a market that has resolved. For
purchase $f$, let $p_f$ denote the price paid per token and $q_f$ the number of
tokens bought. Let $y_f$ denote the token's final payoff: one dollar if the
purchased answer is correct and zero otherwise. The purchase costs $p_fq_f$ and,
if kept until the market resolves, pays $y_fq_f$. Its return is therefore:
\begin{equation}
	r_f
	=
	\frac{y_fq_f-p_fq_f}{p_fq_f}
	=
	\frac{y_f-p_f}{p_f}.
	\label{eq:purchase-return}
\end{equation}

The return in Equation~\eqref{eq:purchase-return} need not be the buyer's
actual profit. The buyer may have sold the tokens before the market resolved.
Instead, it measures what the purchase would have earned if the tokens had
been kept until resolution. We can calculate this return because we observe
both the price paid and the token's final payoff. Unless stated otherwise, this is what we mean by \emph{return} throughout the
paper.\footnote{Returns are not annualized and do not adjust for the time
	between purchase and resolution.} The calculation
excludes trading fees and any rewards that Polymarket pays for leaving orders
available to other participants.

The data record completed purchases and sales but not every way tokens can
enter or leave a wallet. In particular, they do not record all token creation,
redemption, or transfers outside the completed trades in the data. We therefore
cannot reconstruct a complete history of each wallet's holdings from these
transactions. This is another reason why the return in
Equation~\eqref{eq:purchase-return} should not be interpreted as the buyer's
realized profit. The wallet-level profit-and-loss summaries used later in the
paper are separate measures provided with the dataset rather than quantities
we construct from the observed purchases.

The favorite--longshot bias describes a pattern in which low-priced bets (\textit{longshots}) earn
lower returns than high-priced bets (\textit{favorites}). For the main analysis, we classify purchases below 10 cents as longshots and
purchases at or above 90 cents as favorites. All remaining
purchases form the middle-price group. We examine each tail separately.
Specifically, we ask whether longshots lose money on average and whether
favorites earn positive returns. We refer to the pattern as two-sided when
both conditions hold.\footnote{We also repeat the analysis using alternative cutoffs:
		purchases below 5, 15, or 20 cents are classified as longshots, and
		purchases at or above 95, 85, or 80 cents, respectively, are classified
		as favorites. Appendix Table~\ref{tab:thresholds} reports the results.} Table~\ref{tab:price-composition} shows how the purchases used to calculate
returns are divided among the three price groups.

\begin{table}[H]
	\centering
	\caption{Purchases used to calculate returns by price group}
	\label{tab:price-composition}
	\small
	\resizebox{\textwidth}{!}{%
		\begin{tabular}{@{}lrrrrr@{}}
			\toprule
			Price group
			& Price range
			& Purchases
			& Share of purchases
			& Amount paid
			& Share of amount paid \\
			\midrule
			Longshots              & $0<p_f<0.10$        & 100.7M & 18.0\% & $\$273.6$M & 1.2\% \\
			Middle-price purchases & $0.10\leq p_f<0.90$ & 401.5M & 71.6\% & $\$12.56$B & 55.8\% \\
			Favorites              & $0.90\leq p_f\leq1$ & 58.7M  & 10.5\% & $\$9.662$B & 43.0\% \\
			\midrule
			All price groups       & $0<p_f\leq1$         & 560.9M & 100.0\% & $\$22.49$B & 100.0\% \\
			\bottomrule
		\end{tabular}%
	}
	
	\medskip
	\begin{minipage}{0.96\linewidth}
		\footnotesize
		Notes: The table includes only purchases in markets whose final payoff was
		known by March 29, 2026. These purchases form the sample used to calculate
		returns. Amount paid is $\sum_f p_fq_f$ within each price group. Percentages
		may not sum to 100 because of rounding. Authors' calculations from Polymarket
		Users v1.3.
	\end{minipage}
\end{table}

Purchase counts and amounts paid give very different pictures of the two
tails. Longshots account for 18.0 percent of purchases but only 1.2 percent of
the amount paid. Favorites account for 10.5 percent of purchases and 43.0
percent of the amount paid. Large percentage returns on longshots can therefore
correspond to relatively small dollar gains or losses, whereas much smaller
percentage returns on favorites can involve substantially more money.

The preceding table describes purchases and amounts paid. Because one of our
main estimates instead gives each child market the same weight,
Table~\ref{tab:tail-markets-category} reports the number of child markets
represented in each price tail by category.

\begin{table}[H]
	\centering
	\caption{Child markets represented in each price tail, by category}
	\label{tab:tail-markets-category}
	\small
	\begin{tabular*}{\textwidth}{@{\extracolsep{\fill}}lrrr@{}}
		\toprule
		Category
		& \shortstack{Resolved child\\markets}
		& \shortstack{With longshot\\purchases}
		& \shortstack{With favorite\\purchases} \\
		\midrule
		All categories
		& 591,187 (100.0\%)
		& 576,084 (100.0\%)
		& 442,723 (100.0\%) \\
		\midrule
		Politics
		& 24,297 (4.1\%)
		& 23,987 (4.2\%)
		& 23,012 (5.2\%) \\
		Sports
		& 309,782 (52.4\%)
		& 298,445 (51.8\%)
		& 187,101 (42.3\%) \\
		Crypto
		& 205,078 (34.7\%)
		& 201,901 (35.0\%)
		& 185,648 (41.9\%) \\
		Finance
		& 15,147 (2.6\%)
		& 15,006 (2.6\%)
		& 12,782 (2.9\%) \\
		Culture
		& 14,484 (2.4\%)
		& 14,386 (2.5\%)
		& 12,339 (2.8\%) \\
		Tech
		& 3,656 (0.6\%)
		& 3,647 (0.6\%)
		& 3,455 (0.8\%) \\
		Weather
		& 18,743 (3.2\%)
		& 18,712 (3.2\%)
		& 18,386 (4.2\%) \\
		\bottomrule
	\end{tabular*}
	
	\medskip
	\begin{minipage}{0.96\linewidth}
		\footnotesize
		Notes: The table uses the return sample. The first numeric
		column counts distinct child markets that had resolved by
		March 29, 2026. The last two columns count child markets with
		at least one purchase in the indicated price tail. A child
		market can appear in both tail columns. Percentages in
		parentheses report each category's share of the total in
		the corresponding column and may not sum to 100 because
		of rounding.
	\end{minipage}
\end{table}

Sports and Crypto together account for 86.9 percent of the child
markets with longshot purchases and 84.2 percent of those with favorite
purchases. A child market can be represented in both tails because purchases
of its tokens may occur at both low and high prices. The columns therefore
describe the markets represented in each tail rather than dividing child
markets into mutually exclusive groups.

\FloatBarrier

\subsection{Aggregating returns: alternative weighting schemes}
\label{sec:units}

Returns can be aggregated using several weighting schemes. Our main analysis
uses two. The first calculates a return separately for each child market and
then averages those returns, giving every child market the same weight. The
second pools all purchases in each price tail across child markets and
calculates a single return. We refer to it as the pooled return. Because its
denominator is the total amount paid, it is dollar weighted. We report both
because averaging child-market returns and pooling purchases across markets
need not produce the same return.

As a secondary calculation, we pool purchases in each price tail within
each parent event and then average the resulting returns, giving every
represented event equal weight. We treat this measure as secondary because
parent events group questions in different ways: some contain alternative
answers to one question, while others contain related questions that are not
mutually exclusive. Appendix~\ref{app:parent-weighting} provides the formal
definition.

Let $\mathcal{F}$ denote the purchases included in the return sample, and let
$\mathcal{M}$ denote the child markets in which those purchases occurred. Each
purchase $f\in\mathcal{F}$ occurred in one child market, denoted
$m_f\in\mathcal{M}$. For the main analysis, let $T\in\{L,H\}$ index the
longshot and favorite tails, respectively. The corresponding price intervals
are:
\[
\mathcal{I}_L=(0,0.10)
\qquad\text{and}\qquad
\mathcal{I}_H=[0.90,1].
\]
We first sort purchases by child market and price tail. For each child market
$m\in\mathcal{M}$ and tail $T\in\{L,H\}$, let:
\[
\mathcal{F}_{mT}
=
\left\{
f\in\mathcal{F}:m_f=m\text{ and }p_f\in\mathcal{I}_T
\right\}.
\]
This set contains the purchases that occurred in child market $m$ at prices in
tail $T$. Thus, $\mathcal{F}_{mL}$ contains the purchases below 10 cents in
child market $m$, while $\mathcal{F}_{mH}$ contains its purchases at or above
90 cents.

A child market need not have purchases in both tails. Let:
\[
\mathcal{M}_T
=
\left\{
m\in\mathcal{M}:\mathcal{F}_{mT}\neq\varnothing
\right\}
\]
denote the child markets with at least one purchase in tail $T$. Consequently,
$\mathcal{M}_L$ and $\mathcal{M}_H$ need not contain the same child markets,
and they need not be disjoint.

Because the tails classify purchases rather than child markets, a child market
may belong to both $\mathcal{M}_L$ and $\mathcal{M}_H$. For example, in the
child market asking whether Donald Trump would win the 2024 presidential
election, a purchase of a NO token at 8 cents belongs to
$\mathcal{F}_{mL}$, while a purchase of a YES token at 92 cents belongs to
$\mathcal{F}_{mH}$. The two purchases are therefore kept separate when we
calculate the longshot and favorite returns below.

To give each child market the same weight, we first calculate a separate return
for every child market in $\mathcal{M}_T$:
\begin{equation}
	R_{mT}
	=
	\frac{\sum_{f\in\mathcal{F}_{mT}}p_fq_fr_f}
	{\sum_{f\in\mathcal{F}_{mT}}p_fq_f}
	=
	\frac{\sum_{f\in\mathcal{F}_{mT}}q_f(y_f-p_f)}
	{\sum_{f\in\mathcal{F}_{mT}}p_fq_f},
	\qquad
	m\in\mathcal{M}_T,\quad T\in\{L,H\}.
	\label{eq:outcome-roi}
\end{equation}
$R_{mT}$ divides the total terminal gain or loss on purchases in child
market $m$ and tail $T$ by the total amount paid. Purchases within each
child market are therefore dollar weighted; equal weighting applies only
when we average across child markets.

We then average these child-market returns:
\begin{equation}
	R_T^O
	=
	\frac{1}{|\mathcal{M}_T|}
	\sum_{m\in\mathcal{M}_T}R_{mT},
	\qquad
	T\in\{L,H\}.
	\label{eq:equal-outcome}
\end{equation}

Here, $|\mathcal{M}_T|$ is the number of child markets with purchases in
tail $T$. Each market receives equal weight, regardless of the amount
spent on those purchases.

Our next primary aggregation pools purchases across child markets. We first
combine all purchases in tail $T$, then divide their total terminal gain or
loss by the total amount paid:
\begin{equation}
	R_T^D
	=
	\frac{
		\sum_{m\in\mathcal{M}_T}
		\sum_{f\in\mathcal{F}_{mT}}p_fq_fr_f
	}{
		\sum_{m\in\mathcal{M}_T}
		\sum_{f\in\mathcal{F}_{mT}}p_fq_f
	}
	=
	\frac{
		\sum_{m\in\mathcal{M}_T}
		\sum_{f\in\mathcal{F}_{mT}}q_f(y_f-p_f)
	}{
		\sum_{m\in\mathcal{M}_T}
		\sum_{f\in\mathcal{F}_{mT}}p_fq_f
	},
	\qquad
	T\in\{L,H\}.
	\label{eq:roi}
\end{equation}
We call $R_T^D$ the pooled return. It is dollar weighted: equivalently, it is
an average of the purchase returns $r_f$ in which the weight of purchase $f$
is proportional to $p_fq_f$. A child market with twice as much spending
therefore has twice as much influence on the pooled return. Unlike $R_T^O$,
the pooled return does not first assign one return to each child market.

\subsection{Measuring price accuracy}
\label{sec:accuracy}

Returns measure gains and losses relative to the amount paid. We separately
examine whether purchase prices, on average, match final outcomes. Let
$\mathcal{A}\subseteq\mathcal{F}$ denote any nonempty set of purchases. We
calculate:
\begin{equation}
	E(\mathcal{A})
	=
	\frac{\sum_{f\in\mathcal{A}}q_fy_f}
	{\sum_{f\in\mathcal{A}}q_f}
	-
	\frac{\sum_{f\in\mathcal{A}}q_fp_f}
	{\sum_{f\in\mathcal{A}}q_f}
	=
	\frac{\sum_{f\in\mathcal{A}}q_f(y_f-p_f)}
	{\sum_{f\in\mathcal{A}}q_f}.
	\label{eq:probability-error}
\end{equation}

The first term is the share of the total quantity purchased that ultimately
paid one dollar. The second is the quantity-weighted average price paid. We report their difference in percentage
points. A value of zero means that the realized share equals the average
purchase price. A positive value means that the purchased tokens paid one
dollar more often than their prices indicated, while a negative value means
that they did so less often.\footnote{For example, suppose we observe 100
	one-token purchases at 8 cents. The average price matches the realized outcomes
	if eight of those purchases pay one dollar. An individual winning purchase has
	a payoff above its price, but that single realization does not show that the
	8-cent price was inaccurate.}

We aggregate probability errors in parallel with returns. The pooled
probability error combines all purchases in tail $T$ across child markets and
divides their total payoff error by the total number of tokens purchased. It
is therefore token weighted. For equal-child-market averages, we first
calculate the token-weighted probability error within each represented child
market and then average those errors. As a secondary aggregation, we do the
same at the parent-event level and then average across represented events.

Return and probability error therefore answer different questions even when
calculated from the same purchases. At low prices, small probability errors
can generate large percentage gains or losses because the payoff difference
is divided by a small purchase price.

\section{Favorite–Longshot Returns and Price Accuracy}
\label{sec:exists}

\subsection{Returns and probability errors in the two price tails}

Table~\ref{tab:returns_proba_error} reports returns and probability errors in
the two price tails. Panels A and B use our two primary aggregations: equal
weight for each child market and pooled purchases. Panel C gives equal weight
to each parent event and serves as a secondary calculation.\footnote{Appendix
	Table~\ref{tab:thresholds} and Figure~\ref{fig:thresholds} repeat the analysis
	using symmetric 5/95, 15/85, and 20/80 cutoffs. Appendix
	Table~\ref{tab:endpoint} reports the additional restriction excluding endpoint prices.}

\begin{table}[H]
	\centering
	\caption{Returns and probability errors in the longshot and favorite tails}
	\label{tab:returns_proba_error}
	\small
	
	\textit{Panel A. Equal child-market averages}\par\smallskip
	\begin{tabular}{@{}lrrrr@{}}
		\toprule
		Tail
		& Return
		& 95\% interval
		& Probability error
		& 95\% interval \\
		\midrule
		Longshot, $p<0.10$
		& $-6.30\%$
		& $[-8.38,-4.22]\%$
		& $+0.24$ pp
		& $[0.21,0.27]$ pp \\
		Favorite, $p\geq0.90$
		& $+0.277\%$
		& $[0.242,0.312]\%$
		& $+0.271$ pp
		& $[0.237,0.304]$ pp \\
		\bottomrule
	\end{tabular}
	
	\medskip
	\textit{Panel B. Pooled purchases}\par\smallskip
	\begin{tabular}{@{}lrrrr@{}}
		\toprule
		Tail
		& Return
		& 95\% interval
		& Probability error
		& 95\% interval \\
		\midrule
		Longshot, $p<0.10$
		& $-19.35\%$
		& $[-46.99,8.30]\%$
		& $-0.23$ pp
		& $[-0.55,0.09]$ pp \\
		Favorite, $p\geq0.90$
		& $+0.83\%$
		& $[0.60,1.07]\%$
		& $+0.82$ pp
		& $[0.59,1.04]$ pp \\
		\bottomrule
	\end{tabular}
	
	\medskip
	\textit{Panel C. Equal parent-event averages}\par\smallskip
	\begin{tabular}{@{}lrrrr@{}}
		\toprule
		Tail
		& Return
		& 95\% interval
		& Probability error
		& 95\% interval \\
		\midrule
		Longshot, $p<0.10$
		& $+4.09\%$
		& $[1.29,6.90]\%$
		& $+0.20$ pp
		& $[0.15,0.24]$ pp \\
		Favorite, $p\geq0.90$
		& $+0.392\%$
		& $[0.345,0.440]\%$
		& $+0.388$ pp
		& $[0.343,0.432]$ pp \\
		\bottomrule
	\end{tabular}
	
	\medskip
	\begin{minipage}{0.96\linewidth}
		\footnotesize
		Notes: Returns are terminal gains or losses per \textit{dollar paid}. Probability
		errors are terminal gains or losses per \textit{token} and are reported in percentage
		points. Panel A averages child markets equally, Panel B pools purchases, and
		Panel C averages parent events equally. Intervals in Panels A and B are
		clustered by parent event; Panel C uses standard errors across events.
	\end{minipage}
\end{table}

When child markets receive equal weight, returns display the classic
favorite--longshot pattern. Longshot purchases lose $6.30$ percent on average,
with a 95 percent confidence interval from $-8.38$ to $-4.22$ percent. Favorite
purchases earn $0.277$ percent on average, with an interval from $0.242$ to
$0.312$ percent. In turn, probability errors tell a different story for
longshots. Despite their negative average return, their average probability
error is $+0.24$ percentage points. The corresponding average for favorites is
$+0.271$ percentage points.

The signs for longshots diverge only after markets are averaged. Within a given
market, return and probability error necessarily have the same sign. Converting
a probability error into a return, however, requires dividing it by the average
purchase price. An error of a given size therefore produces a larger percentage
gain or loss in a lower-priced market. In these data, that scaling magnifies the
negative errors enough to make the average return negative even though the
average probability error is positive.

Pooling purchases across child markets likewise produces negative longshot
returns and positive favorite returns. Longshots lose $19.35$ percent per
dollar paid. Although this point estimate is economically large, the confidence
interval is quite wide and includes zero, as it ranges from $-46.99$ to $+8.30$
percent. Favorites earn $+0.83$ percent, with an interval from $0.60$ to $1.07$
percent. In the pooled data, probability errors have the same signs as returns:
the longshot error is $-0.23$ percentage points, while the favorite error is
$+0.82$ percentage points. This agreement is mechanical, both measures use the
same total gain or loss in the numerator, but return divides by dollars paid
whereas probability error divides by tokens purchased.

The picture changes when parent events receive equal weight. The average
longshot return becomes $+4.09$ percent and the average probability error is
$+0.20$ percentage points; both favorite measures remain positive. This
calculation first pools purchases across child markets within each parent event
and then gives every parent event equal weight. An event with many child markets
therefore receives no more weight than an event with one. The reversal in the
longshot result shows why the choice of aggregation unit matters.

\subsection{Why does the longshot result change?}
\label{sec:aggregation-change}

The longshot return is negative when child markets receive equal weight and
when purchases are pooled, but positive when parent events receive equal
weight. What accounts for this reversal?

Table~\ref{tab:decomposition} decomposes the differences among these estimates
by changing one set of weights at a time.\footnote{We focus on longshots because
	favorite returns remain positive in all four calculations.}
	
\begin{table}[H]
	\centering
	\caption{Decomposing changes in longshot returns}
	\label{tab:decomposition}
	\small
	\begin{tabular}{@{}>{\raggedright\arraybackslash}p{0.14\linewidth}>{\raggedright\arraybackslash}p{0.58\linewidth}>{\raggedleft\arraybackslash}p{0.16\linewidth}@{}}
		\toprule
		\multicolumn{3}{@{}l}{\textit{Panel A. Longshot returns}}\\
		\addlinespace[0.25em]
		\multicolumn{2}{@{}l}{Calculation}
		& Longshot return\\
		\midrule
		\multicolumn{2}{@{}l}{1. Equal child-market weights}
		& $-6.300\%$\\
		\addlinespace[0.25em]
		\multicolumn{2}{@{}l}{2. Equal parent and child-market weights}
		& $+4.029\%$\\
		\addlinespace[0.25em]
		\multicolumn{2}{@{}l}{3. Equal parent weights; purchases pooled}
		& $+4.092\%$\\
		\addlinespace[0.25em]
		\multicolumn{2}{@{}l}{4. All purchases pooled}
		& $-19.348\%$\\
		\addlinespace[0.75em]
		\multicolumn{3}{@{}l}{\textit{Panel B. Changes between calculations}}\\
		\addlinespace[0.25em]
		Comparison
		& What changes
		& Return change\\
		\midrule
		1 to 2
		& Give every parent event equal weight
		& $+10.329$ pp\\
		\addlinespace[0.25em]
		2 to 3
		& Pool purchases within each parent event
		& $+0.063$ pp\\
		\addlinespace[0.25em]
		3 to 4
		& Weight parent events by longshot purchase dollars
		& $-23.440$ pp\\
		\bottomrule
	\end{tabular}
	
	\medskip
	\begin{minipage}{0.94\linewidth}
		\footnotesize
		Notes: Longshots have purchase prices below 10 cents. Panel B reports the
		change in longshot return between consecutive calculations in Panel A. A
		positive value denotes an increase. Calculations 1 and 4 are the two primary
		calculations, calculation 3 is the secondary calculation, and calculation 2
		is used only to separate the two changes between calculations 1 and 3.
		``pp'' denotes percentage points.
	\end{minipage}
\end{table}

Moving from calculation 1 to calculation 2 raises the longshot return by
$10.329$ percentage points, from $-6.300$ to $+4.029$ percent. Parent events
with more child markets therefore tend to have lower longshot returns and pull
down the equal-child-market average, in which each child market counts
separately.

By contrast, moving from calculation 2 to calculation 3 raises the return by
only $0.063$ percentage points, to $+4.092$ percent. Whether child markets
within a parent event are weighted equally or according to spending on
longshots makes almost no difference. The reversal from the equal-child-market
estimate to the equal-parent-event estimate is therefore driven almost
entirely by the number of child markets listed under each parent event.

The final transition replaces equal weights across parent events with weights
proportional to the amount spent on longshots in each event. This lowers the
return by $23.440$ percentage points, from $+4.092$ to $-19.348$ percent.
Parent events that attract more spending on longshots tend to have lower
returns. Thus, the average parent event has a positive longshot return, while
the dollars spent on longshots earn a negative return because spending is
concentrated in lower-return events.

The results isolate two distinct sources that explain the differences among the
estimates. Parent events with many child markets pull down the
equal-child-market return, while parent events with more spending on longshots
pull down the pooled return. The way purchases are weighted within parent
events contributes almost nothing. Appendix
Section~\ref{app:aggregation-decomposition} presents the formal
identities.\footnote{Appendix Table~\ref{tab:menusize} reports longshot returns
	by the number of child markets within each parent event. These returns do not
	fall steadily as the number of child markets increases.}

\subsection{Returns and probability errors across purchase prices}

\begin{figure}[ht!]
	\caption{Returns and probability errors across purchase prices, giving
		each child market equal weight}
	\label{fig:full-price-equal-market}
	\centering
	\includegraphics[width=0.94\linewidth]
	{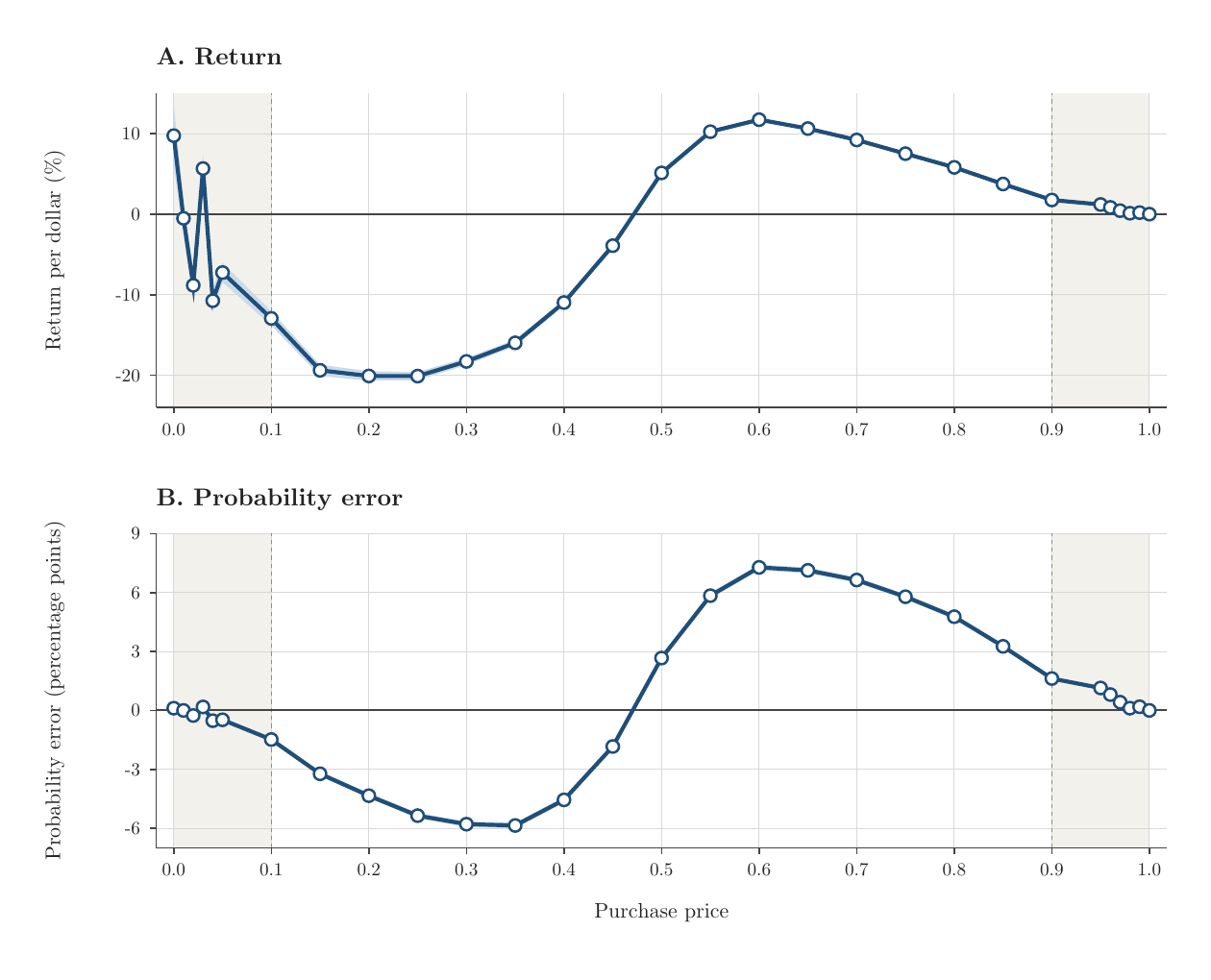}
	
	\begin{minipage}{0.94\linewidth}
		\footnotesize
		Notes: Each point represents a purchase-price bin. Within each bin,
		the measures are first calculated separately for each represented
		child market and then averaged, so every child market receives the
		same weight. Return is terminal gain or loss per \textit{dollar paid}.
		Probability error is terminal gain or loss per \textit{token} and is reported
		in percentage points. Outer bins have width 0.01 and interior bins
		have width 0.05. Light shading marks the longshot and favorite tails.
		Ribbons are 95 percent confidence intervals clustered by parent event.
	\end{minipage}
\end{figure}

Figure~\ref{fig:full-price-equal-market} traces returns and probability errors across the full price range. It gives every represented child market the same weight within each price bin. Returns are negative across most of the lower half of the price range and positive across the upper half. Losses are largest around 20--25 cents, while gains peak around 60 cents and then decline toward zero as prices approach one dollar. Probability errors follow the same broad pattern.

Returns fluctuate more sharply at the far left because low purchase prices
magnify modest positive or negative probability errors. We can therefore
conclude that the tail estimates are part of a broader relationship with
purchase price rather than an artifact of the 10/90 cutoffs.

\FloatBarrier

\section{How Does the Bias Vary Across categories?}
\label{sec:generality}

The aggregate results combine markets from seven different categories. We now ask
whether each category displays the same two-sided pattern: negative longshot
returns and positive favorite returns. We repeat the tail calculations
separately by category using the two primary aggregation methods. We then examine
the full price curves within each category.

\subsection{Longshot and favorite returns by category}

Table~\ref{tab:category-returns} reports longshot and favorite returns
separately by category. Panel A gives every represented child market equal weight,
while Panel B pools purchases within each category.

\begin{table}[H]
	\centering
	\caption{Longshot and favorite returns by market category}
	\label{tab:category-returns}
	\small
	\begin{tabular*}{\textwidth}{@{\extracolsep{\fill}}lrrrr@{}}
		\toprule
		\multicolumn{5}{@{}l}{\textit{Panel A. Equal child-market averages}}\\
		\addlinespace[0.25em]
		category
		& Longshot return
		& 95\% interval
		& Favorite return
		& 95\% interval\\
		\midrule
		Crypto   & $-14.84\%$ & $[-17.81,-11.87]\%$ & $+0.643\%$ & $[0.589,0.697]\%$\\
		Politics & $-16.34\%$ & $[-23.30,-9.39]\%$  & $+1.044\%$ & $[0.904,1.184]\%$\\
		Sports   & $+2.43\%$  & $[-0.93,5.79]\%$    & $-0.230\%$ & $[-0.285,-0.174]\%$\\
		Finance  & $-2.62\%$  & $[-11.45,6.22]\%$   & $+0.085\%$ & $[-0.182,0.353]\%$\\
		Weather  & $-25.15\%$ & $[-32.30,-18.00]\%$ & $+0.502\%$ & $[0.419,0.585]\%$\\
		Culture  & $-26.49\%$ & $[-35.84,-17.14]\%$ & $+0.735\%$ & $[0.515,0.954]\%$\\
		Tech     & $-20.68\%$ & $[-35.74,-5.63]\%$  & $+0.816\%$ & $[0.459,1.174]\%$\\
		\addlinespace[0.75em]
		\multicolumn{5}{@{}l}{\textit{Panel B. Pooled purchases}}\\
		\addlinespace[0.25em]
		category
		& Longshot return
		& 95\% interval
		& Favorite return
		& 95\% interval\\
		\midrule
		Crypto   & $-12.63\%$ & $[-19.84,-5.41]\%$  & $+0.537\%$ & $[0.329,0.745]\%$\\
		Politics & $-46.15\%$ & $[-68.34,-23.96]\%$ & $+1.419\%$ & $[0.857,1.982]\%$\\
		Sports   & $+18.11\%$ & $[-74.68,110.91]\%$ & $+0.137\%$ & $[-0.100,0.373]\%$\\
		Finance  & $-72.40\%$ & $[-93.32,-51.48]\%$ & $+1.826\%$ & $[1.187,2.465]\%$\\
		Weather  & $+24.74\%$ & $[2.49,47.00]\%$    & $+0.031\%$ & $[-0.296,0.358]\%$\\
		Culture  & $-18.78\%$ & $[-43.32,5.76]\%$   & $+0.829\%$ & $[0.351,1.308]\%$\\
		Tech     & $-18.91\%$ & $[-42.29,4.46]\%$   & $+0.846\%$ & $[0.227,1.464]\%$\\
		\bottomrule
	\end{tabular*}

	\medskip
	\begin{minipage}{0.96\linewidth}
		\footnotesize
		Notes: Longshots have purchase prices below 10 cents and favorites have
		prices at or above 90 cents. Panel A first calculates returns within each
		child market and then averages child markets equally. Panel B pools
		purchases within each category and weights them by the amount paid.
		Intervals are clustered by parent event.
	\end{minipage}
\end{table}

Table~\ref{tab:category-returns} shows substantial differences across categories.
Crypto and Politics provide the clearest evidence of a two-sided
favorite--longshot bias. In Crypto, longshots lose 14.84 percent when child
markets receive equal weight and 12.63 percent when purchases are pooled. In
Politics, the corresponding losses are 16.34 and 46.15 percent. Favorite
returns are positive in both categories under both aggregation methods. Thus, for Crypto
and Politics, both tails have the classic signs and all four confidence
intervals exclude zero.

Culture and Tech also have negative longshot and positive favorite point
estimates under both methods. Their equal-child-market estimates are clearly
different from zero, but their pooled longshot intervals include it. These
categories therefore display the two-sided pattern more clearly across listed
markets than across purchase dollars.

Finance depends strongly on aggregation. When purchases are pooled, longshots
lose $72.40$ percent and favorites earn $1.826$ percent; both confidence
intervals exclude zero. Under equal-child-market weighting, both estimates are
much closer to zero and both confidence intervals include zero. The Finance
pattern is therefore concentrated in the dollars invested rather than broadly
distributed across listed markets.

Weather changes direction across the two calculations. With equal weight on
each child market, longshots lose $25.15$ percent and favorites earn $0.502$
percent. When purchases are pooled, however, longshots earn $24.74$ percent. Sports is the clearest
counterexample: its longshot returns are positive under both methods, while
its favorite result is not consistently positive.

The favorite--longshot bias is therefore not universal across Polymarket. It
is most robust in Crypto and Politics, appears under some calculations in
Culture, Tech, Finance, and Weather, and is absent in Sports. category and
aggregation method jointly determine where the pattern appears.

\subsection{Prices and realized outcomes by category}

\begin{figure}[ht!]
	\caption{Transaction prices and realized outcomes by market category}
	\label{fig:categorycalibration}
	\centering
	\includegraphics[width=0.88\linewidth]{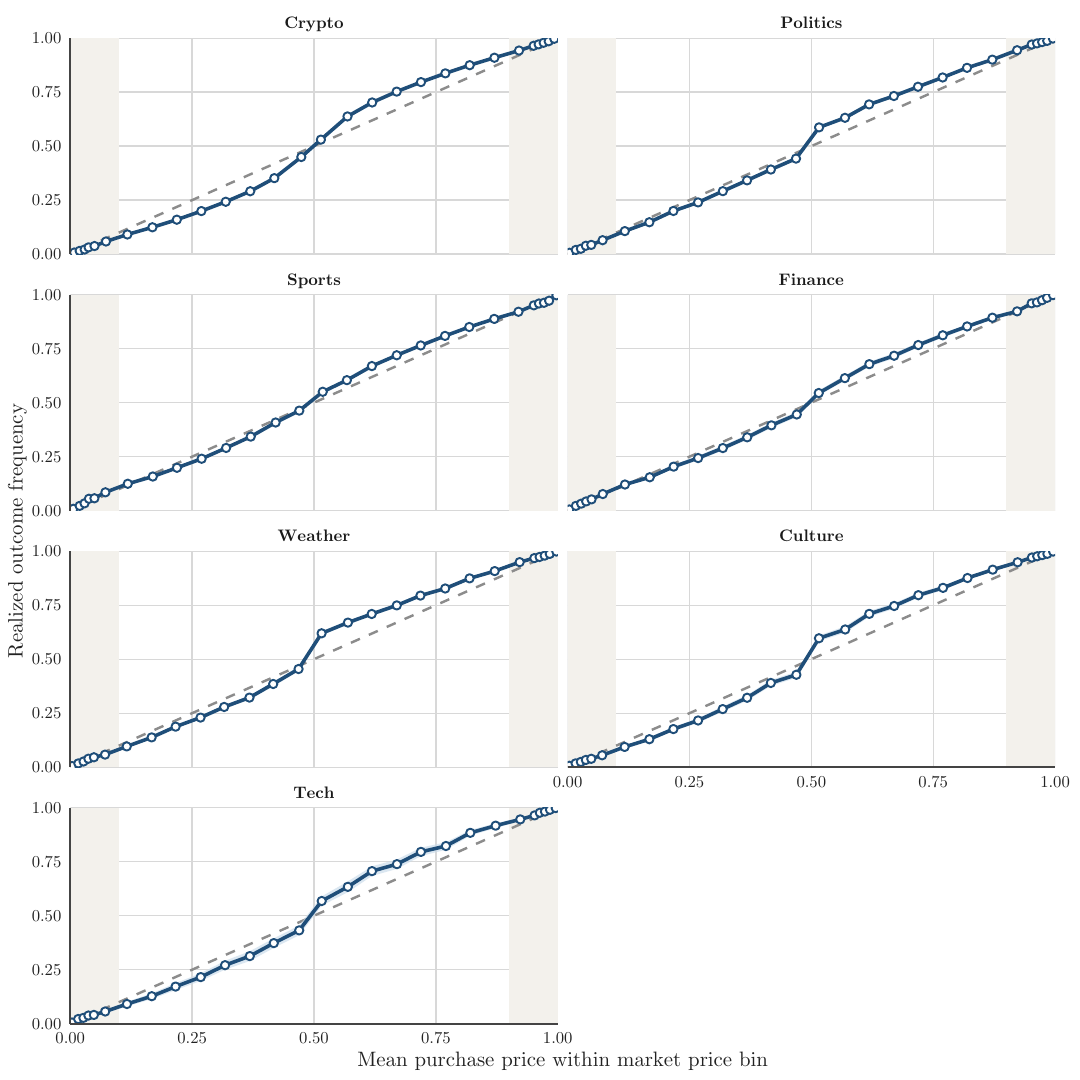}
	
	\begin{minipage}{0.94\linewidth}
		\footnotesize
		Notes: Each point gives equal weight to listed-market purchase-price-bin
		cells within category. Coordinates are the mean purchase price and realized
		outcome frequency. The dashed line marks equality between the average price
		and the realized outcome frequency, and the shaded regions mark
		$p<0.10$ and $p\geq0.90$. Ribbons are 95 percent confidence intervals
		clustered by parent event. Cells with fewer than 50 markets are omitted. A
		market can enter multiple bins, so the curves do not algebraically reproduce
		the tail estimates.
	\end{minipage}
\end{figure}

Figure~\ref{fig:categorycalibration} extends the price-accuracy analysis in Figure~\ref{fig:full-price-equal-market} by showing the relationship separately for each category. Within each category and price bin, it compares the average purchase price with how often the purchased tokens pay one dollar, giving equal weight to each child-market in each price-bin cell. Points on the dashed line indicate that tokens pay one dollar as often as their prices imply; points below or above the line indicate that they pay less or more often, respectively. While Table~\ref{tab:category-returns} summarizes returns in the longshot and favorite tails, Figure~\ref{fig:categorycalibration} shows whether the differences across categories extend throughout the price range.

Most categories display the same broad shape as the platform-wide curve: realized
outcome frequencies tend to lie below prices through much of the lower price
range and above prices through much of the upper range. Sports remains closer
to the equality line, consistent with its failure to display the classic
two-sided pattern in the tail estimates. The category differences are therefore
not confined to a single pooled return calculation.

The patterns in Figure~\ref{fig:categorycalibration} need not match the tail
returns in Table~\ref{tab:category-returns}. In Panel A of the table, each
child market contributes one return to each tail in which it appears. In the
figure, the same child market contributes separately to every price bin in
which it has purchases and may therefore affect several points on the curve.
Table~\ref{tab:category-returns} therefore identifies the categories in which the two-sided
return pattern appears in the tails, while
Figure~\ref{fig:categorycalibration} shows how prices and realized outcomes
relate across the full price range within each category.

\FloatBarrier
\clearpage
\section{Trading Conditions and Tail Returns}
\label{sec:explanations}

We next examine how tail returns vary with trading conditions.
Prediction markets can become more accurate as they absorb information
\citep{wolferszitzewitz2004}. We therefore begin by asking whether
longshot losses persist in markets with substantial previous trading.

We then ask whether longshot buyers lose more when they accept a
seller's price than when they post their own offer and wait. Buying
immediately can mean paying more: the buyer accepts the seller's
asking price instead of offering less and waiting for a seller to
agree. \citet{akey2026} actually find that Polymarket traders who accept
existing orders tend to perform worse than those who post them,
and their calculations suggest that the cost of accepting existing
offers contributes to some traders' losses. Such costs could be
particularly important for longshots. Paying one cent more for a
token is a large increase on a purchase price of a few cents, but
a small increase on a price close to one dollar.

Posting an offer to buy carries a different risk: a seller may accept it
after bad news has made the posted price too high
\citep{foucault1999}. Such over-payments could contribute to the
longshot losses measured at resolution. For purchases made through
posted offers, we therefore look for early price declines that
would be consistent with this explanation.

We also consider whether fees and cash requirements make trading
against overpricing less attractive \citep{shleifervishny1997}.
Although our return estimates exclude fees, fees reduce the expected profit
from selling an overpriced token. If they discourage these sales,
overpricing may persist and longshot buyers may lose more, even
before fees. We therefore compare markets with and without
reported fees, the comparison uses a market-level fee indicator,
not the fee paid on each transaction.

Finally, we compare events with and without shared collateral.
Shared collateral reduces the cash required to create tokens for
sale across mutually exclusive outcomes. If committing cash
prevents traders from correcting overpricing, longshot losses
should be smaller where collateral can be shared.

\begin{table}[!t]
	\centering
	\caption{Tail returns under different trading conditions}
	\label{tab:tradingconditions}
	\small
	\setlength{\tabcolsep}{5pt}
	\begin{tabular}{@{}lrrrr@{}}
		\toprule
		& \multicolumn{2}{c}{Equal child markets}
		& \multicolumn{2}{c}{Pooled purchases}\\
		\cmidrule(lr){2-3}\cmidrule(l){4-5}
		& Longshots & Favorites & Longshots & Favorites\\
		\midrule
		\multicolumn{5}{l}{\textit{Panel A. Previous trading}}\\
		Least previous trading & --- & --- & $-4.08\%$ & $+0.491\%$\\
		Second group & --- & --- & $-6.53\%$ & $+0.384\%$\\
		Third group & --- & --- & $-4.15\%$ & $+0.323\%$\\
		Most previous trading & --- & --- & $-28.72\%$ & $+1.265\%$\\
		\addlinespace
		\multicolumn{5}{l}{\textit{Panel B. Buyer action}}\\
		Posted purchase offer
		& $-3.50\%$ & $+0.641\%$ & $-17.56\%$ & $+0.816\%$\\
		Accepted seller's offer
		& $-31.17\%$ & $-0.460\%$ & $-22.17\%$ & $+0.877\%$\\
		\addlinespace
		\multicolumn{5}{l}{\textit{Panel C. Reported fee}}\\
		No fee flag
		& $-5.20\%$ & $+0.260\%$ & $-20.78\%$ & $+0.864\%$\\
		Fee flag
		& $-13.03\%$ & $+0.371\%$ & $+10.03\%$ & $+0.333\%$\\
		\addlinespace
		\multicolumn{5}{l}{\textit{Panel D. Shared collateral}}\\
		No
		& $-5.94\%$ & $+0.182\%$ & $-9.79\%$ & $+0.588\%$\\
		Yes
		& $-7.80\%$ & $+0.644\%$ & $-25.76\%$ & $+1.286\%$\\
		\bottomrule
	\end{tabular}
	
	\medskip
	\begin{minipage}{0.96\linewidth}
		\footnotesize
		Notes: Longshots have purchase prices below 10 cents and favorites
		at or above 90 cents. Returns compare terminal payoffs with purchase
		costs before fees. Panel A ranks entries defined by child market,
		purchase month, and price bin according to the dollars traded in
		that child market in earlier months. Rankings are formed separately
		within category, month, and tail; only pooled returns are reported.
		Panel B distinguishes buyers who post a purchase offer from buyers
		who accept a seller's existing offer. Panel C uses the reported fee
		indicator, not the fee paid on individual transactions. Panel D uses
		verified metadata on shared collateral.
		Appendix~\ref{app:tradingconditions} reports the available confidence
		intervals, sample details, and outstanding checks. Dashes denote
		calculations not included in the existing output, not zeros.
	\end{minipage}
\end{table}

\subsection{Previous trading}

We group purchases according to the amount previously traded in
their child market. We do this separately for longshots and favorites
within each category and purchase month. Each child market contributes
one entry for each price bin in which purchases occurred. We measure
previous trading as the total dollars paid for YES and NO tokens
in that child market in earlier months, across all prices. We rank
the entries by this amount and divide them into four approximately
equal groups, from least to most previous trading.\footnote{For
	illustration, return to the child market ``Will Donald Trump win
	the 2024 presidential election?'' Suppose its only longshot purchases
	in October 2024 were of NO tokens at 4 and 8 cents. This market
	contributes two entries to the Politics longshot ranking for that
	month. Both receive the same previous-trading amount: the total
	dollars paid for YES and NO tokens in that market, at any price,
	before October 2024. If this ranking contains eight entries
	altogether, each group contains two. Purchases assigned to group 4
	are then pooled with group 4 longshot purchases from every other
	category and month.}

For each tail, we pool purchases assigned to the same group across
all categories and months. Each group's return is its total terminal
gain or loss divided by the total amount paid. This gives four
longshot returns and four favorite returns.

Longshot returns are negative in all four groups, while favorite
returns are positive throughout
(Table~\ref{tab:tradingconditions}, Panel A).
The largest longshot loss and favorite gain occur in the group
with the most previous trading: $-28.72$ percent and $+1.26$
percent, respectively. The losses are therefore not confined to
markets where little trading has occurred.

Higher trading volume may reflect participation by more bettors.
The stronger pattern in the most heavily traded group is consistent
with the model of \citet{ottavianisorensen2010}. In their model, the
same share of bets backing a favorite is a more reliable indication
that it will win when those bets come from more independently informed
bettors. But the odds depend on the share of money bet, regardless of
how many bettors are behind it. As participation increases, favorites
can therefore become more likely to win than their odds suggest, and
longshots less likely, strengthening the favorite--longshot bias.

\subsection{How purchases are made}

With purchases pooled, longshot losses and favorite gains occur
under both purchase methods
(Table~\ref{tab:tradingconditions}, Panel B).
Longshot purchases lose $17.56$ percent when the buyer posted the
offer and $22.17$ percent when the buyer accepted the seller's
offer. Favorite purchases earn $0.82$ and $0.88$ percent,
respectively. The longshot confidence interval includes zero
for buyers accepting a seller's offer but excludes zero for
buyers posting the offer; both favorite intervals exclude zero.

When child markets receive equal weight, longshot returns are
$-3.50$ percent for posted offers and $-31.17$ percent for accepted
offers. The difference between the two methods is therefore much
larger under equal child-market weighting than in pooled purchases.
Favorite returns are $+0.64$ percent when the buyer posted the offer
but $-0.46$ percent when the buyer accepted the seller's offer.
Under this weighting, buyers accepting sellers' offers lose money
at both tails, rather than exhibiting the combination of longshot
losses and favorite gains.

Longshot losses are larger among buyers accepting sellers' prices
under either weighting scheme. But they also occur when buyers
post their own offers, so they are not confined to purchases made
at sellers' asking prices.

To investigate whether buying at outdated prices could help explain
the losses on posted-offer purchases, we ask whether transaction
prices fall below the purchase prices soon afterward. For this
calculation, we combine purchases of the same token made through
posted offers during fixed five-minute periods, such as
12:00--12:05 and 12:05--12:10. Within each period, we divide the total
amount paid by the number of tokens purchased to obtain the average
purchase price. We classify these purchases as longshots if that
average is below 10 cents and as favorites if it is at least 90 cents.

We then value the purchased tokens at transaction prices observed
approximately five minutes, one hour, and one day later.\footnote{Time
	is measured from the start of the five-minute purchase period.
	For each horizon, we select the first five-minute price interval
	starting at or after the target time and use its last transaction
	price. We allow the selected interval to start up to 10 minutes
	after the five-minute target, one hour after the one-hour target,
	and six hours after the one-day target. The target must precede
	resolution, and the selected price must be strictly between zero
	and one dollar. Purchases without a qualifying later price are
	excluded from that horizon's comparison.}
For longshots and favorites separately, we add up the value of
the tokens at these later prices, subtract their total purchase
cost, and divide the difference by that cost.

\begin{figure}[H]
	\centering
	\caption{Returns after purchases made through posted purchase offers}
	\label{fig:postpurchaseprices}
	\includegraphics[width=0.97\linewidth]{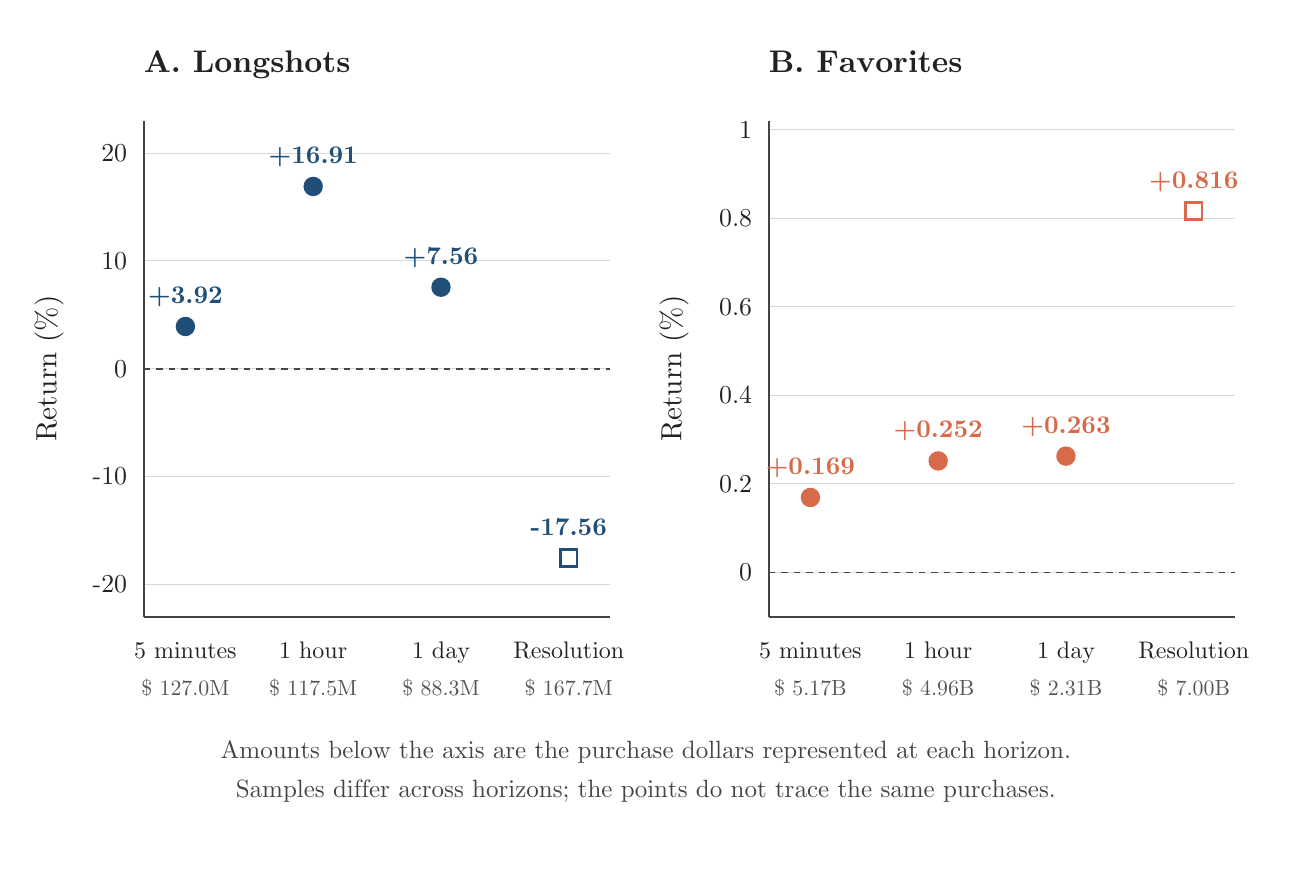}
	\begin{minipage}{0.95\linewidth}
		\footnotesize
		Notes: Short-horizon returns value purchased tokens at the closing
		transaction price of the first five-minute interval satisfying the
		matching rule described in the text. The amounts below the axis are
		the purchase dollars represented at each point. Purchases without
		a qualifying later price are omitted from that horizon, so the
		points do not follow a fixed set of purchases. The resolution
		points instead use final payouts on all resolved tail purchases
		made through posted offers. For the short-horizon calculations,
		purchases of the same token within a five-minute period are
		classified together using their average purchase price; for the
		terminal calculation, each purchase is classified using its own
		price. Panels use different vertical scales. No confidence
		intervals are available for the horizon-specific short-horizon
		estimates; the common-sample estimates and parent-event-clustered
		intervals are reported below and in Appendix
		Table~\ref{tab:postpurchasebalanced}.
		Appendix Table~\ref{tab:postpurchase} reports the amounts represented
		and matching windows.
	\end{minipage}
\end{figure}

The resulting returns are positive at all three horizons
(Figure~\ref{fig:postpurchaseprices}). Longshot returns are $+3.92$,
$+16.91$, and $+7.56$ percent at approximately five minutes, one
hour, and one day, respectively. Favorite returns are $+0.169$,
$+0.252$, and $+0.263$ percent. Thus, in these comparisons,
subsequent transaction prices value the purchases above what
buyers paid. We do not observe the early price declines that
would support the outdated-offer explanation for terminal
longshot losses.\footnote{These positive returns do not overturn the longshot losses
	calculated from final payouts: subsequent transaction prices can
	also be too high relative to the tokens' chances of winning.
	Nor do they establish when the terminal losses arise. A set of
	purchases enters a short-horizon comparison only if we find a
	qualifying later price, whereas the terminal calculation does
	not require one. The points therefore describe different sets
	of purchases, not the successive returns on the same purchases. Finally, these are valuations at observed transaction prices,
	not profits from actual sales. A later buyer may pay a seller's
	asking price that the original buyer could not obtain by selling
	immediately. The positive estimates therefore do not rule out
	purchases at outdated prices.}

Restricting the calculation to the same purchase groups observed at all three
interim horizons and at resolution does not change the interpretation. For
longshots, returns are $+4.85$, $+6.96$, and $+6.62$ percent after five
minutes, one hour, and one day, before falling to $-36.27$ percent at
resolution. The corresponding 95 percent confidence intervals, clustered by
parent event, are $[4.08,5.62]$, $[5.46,8.46]$, $[2.14,11.10]$, and
$[-60.88,-11.66]$ percent. For favorites, the common-sample returns are
$+0.017$, $+0.014$, $+0.308$, and $+1.939$ percent, with intervals
$[-0.003,0.037]$, $[-0.020,0.047]$, $[0.187,0.429]$, and
$[1.378,2.499]$ percent. The common sample contains 6,526,943 longshot entry
groups representing \$64.62 million and 1,179,457 favorite entry groups
representing \$1.710 billion. These account for 41.3 and 20.7 percent of all
entry groups and 37.7 and 25.6 percent of all entry-group purchase dollars,
respectively. Appendix Table~\ref{tab:postpurchasebalanced} reports the full
coverage calculation.

\subsection{Reported fees and shared collateral}

Longshot losses and favorite gains are present in markets with
no reported trading fee
(Table~\ref{tab:tradingconditions}, Panel C).
Giving each child market equal weight, longshot purchases return
$-5.20$ percent and favorite purchases return $+0.260$ percent.
The corresponding pooled returns are $-20.78$ and $+0.864$ percent,
although the pooled longshot confidence interval includes zero.

Markets reporting fees do not consistently have larger longshot
losses. Their equal-child-market longshot return is more negative,
at $-13.03$ percent, but their pooled longshot return is positive,
at $+10.03$ percent. Favorite returns are $+0.371$ percent with
equal child-market weights and $+0.333$ percent with purchases
pooled. Both pooled confidence intervals in this fee-reporting
group include zero. The comparison therefore does not show consistently larger
longshot losses in markets reporting fees, as the proposed
explanation would suggest. It also establishes that the
combination of longshot losses and favorite gains is not confined
to those markets.

Shared collateral concerns the cash needed to create tokens
for sale. Consider an election with three candidates, exactly
one of whom will win. Without shared collateral, one dollar
creates a YES and a NO token for one candidate. Creating all
three pairs before selling any tokens requires three dollars.
With shared collateral, one dollar can instead support a set
containing one YES token for each candidate, because exactly
one of those tokens will pay one dollar.\footnote{One dollar
	first creates a YES and a NO token for candidate A. The NO
	token for A can then be converted into YES tokens for B and C,
	leaving one YES token for each candidate. This example abstracts
	from conversion fees. The conversion rule is described in
	\href{https://github.com/Polymarket/neg-risk-ctf-adapter}
	{Polymarket's contract documentation}.}
If the cash required to create tokens limits selling at
overpriced levels, sharing collateral could make those sales
easier and reduce longshot losses.

The estimates go in the opposite direction
(Table~\ref{tab:tradingconditions}, Panel D).
Longshot returns are $-7.80$ percent with shared collateral
and $-5.94$ percent without it when child markets receive
equal weight. With purchases pooled, they are $-25.76$ and
$-9.79$ percent, respectively. Favorite returns with shared
collateral are $+0.64$ percent under equal child-market weighting
and $+1.29$ percent in pooled purchases.
Without shared collateral, favorite returns are $+0.182$ percent under equal
child-market weighting and $+0.588$ percent in pooled purchases. Thus favorite
returns are positive under both collateral regimes, but larger in the events
where collateral can be shared. As with the longshot comparison, this is a
cross-sectional difference rather than a causal estimate.

These results do not support the prediction that longshot losses
are smaller where collateral can be shared. They do not establish
that sharing collateral increases losses either: we compare
different events, not the same events before and after a change
in collateral requirements. We also observe whether collateral
can be shared, not the cash each trader actually commits.
Appendix Section~\ref{app:collateral} reports the metadata coverage
and the transactions that cannot be reconstructed.

Taken together, the comparisons show that longshot losses persist
after substantial previous trading, among buyers who post their
own offers, in markets without reported fees, and where collateral
can be shared. The short-horizon comparisons also do not show
early price declines after posted-offer purchases. Pooled purchases
retain longshot losses and favorite gains across previous-trading
groups and both purchase methods, but this two-sided pattern is
not universal under equal child-market weighting. The section
thus narrows the circumstances to which the losses can be confined
without identifying a single cause of the favorite--longshot bias.

\FloatBarrier
\clearpage

\section{Who Buys Longshots and Favorites?}
\label{sec:traders}

In this Section we ask whether some wallets repeatedly buy longshots or favorites.
We then compare their shares of tail purchases with their shares of longshot
losses and favorite gains. We also assess whether longshot losses are confined
to inexperienced wallets.

\subsection{Classifying wallets by past tail demand}

For each month, we classify wallets using purchases made during the preceding
six calendar months. We only include wallets that made purchases in at least two of
those months and pass the sample screens described in
Section~\ref{sec:data}. We calculate the share of their purchase dollars spent
below 10 cents and, separately, the share spent at or above 90 cents.

Because these shares vary with trading activity and category choice, we estimate
two separate regressions for each month. The first explains the longshot share
using total purchase volume, active months, market participation, and the
wallet's Politics, Sports, and Crypto shares of trading volume. The second
uses the same predictors to explain the favorite share.

Within each
month, we rank wallets by how much their observed tail share exceeds the share
predicted by these characteristics. The highest longshot decile forms the
longshot group, and the highest favorite decile forms the favorite group. The
procedures are identical except for the price range used to calculate the
purchase share. A wallet can enter or leave
either group over time. Neither classification uses purchases or outcomes from
the following month.
Appendix~\ref{app:traders} provides the full specification and alternative
classifications.

\subsection{Does past tail demand predict future demand?}

The classifications do predict where wallets trade in the following month.
Although each group contains 10 percent of classified wallets, the longshot
group supplies 26.6 percent of longshot purchase dollars and the favorite
group supplies 15.1 percent of favorite purchase dollars.

\begin{figure}[H]
	\caption{Persistence of longshot and favorite demand}
	\label{fig:taildemandpersistence}
\centering
\includegraphics[width=0.94\linewidth]{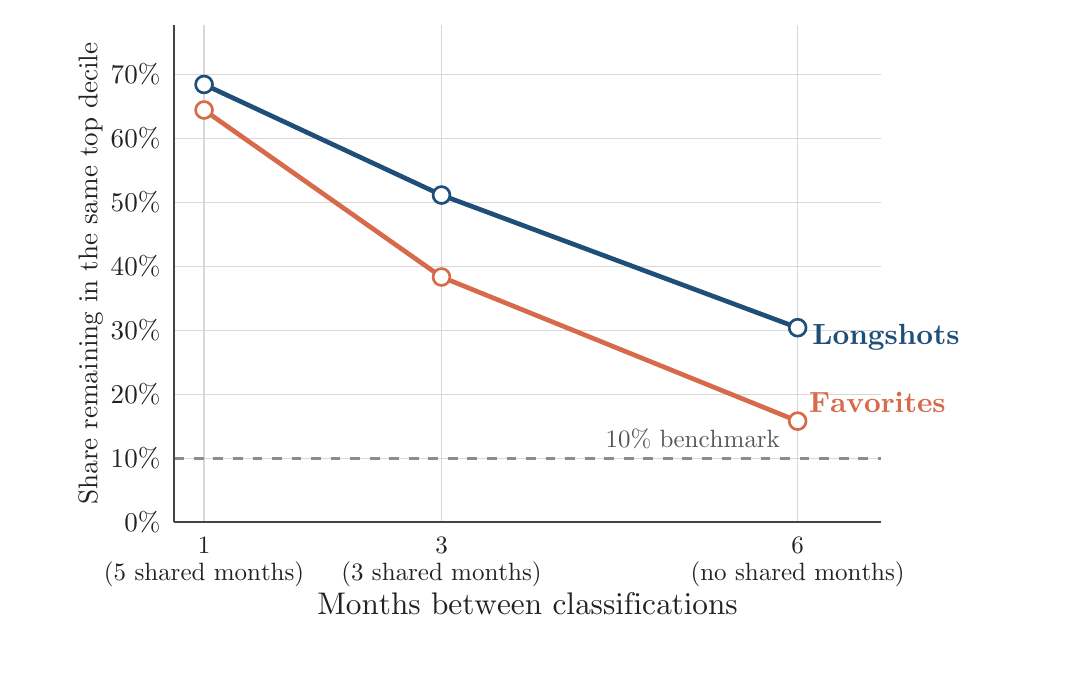}
\begin{minipage}{0.95\linewidth}
\footnotesize
Notes: Among wallets classified again at the stated horizon, the figure reports
the share of each initial top decile that is again in the corresponding top
decile. Classifications one month apart share five months of purchase history;
classifications three months apart share three months; and classifications six
months apart share none. The dashed line marks the 10 percent benchmark.
\end{minipage}
\end{figure}

The classifications remain predictive at longer horizons. Figure~\ref{fig:taildemandpersistence} follows wallets that begin in one of
the two top deciles. After one month, 68.5 percent of the longshot group and
64.5 percent of the favorite group remain
in the corresponding top decile. These comparisons partly reuse the same
purchase history: classifications one month apart share five of their six
lookback months, and classifications three months apart share three. Six
months apart, the histories no longer overlap. But even then, 30.4 percent of the
original longshot group and 15.8 percent of the original favorite group remain
in the same top decile. Past demand therefore predicts future demand at both
tails, with stronger persistence in the longshot group.

\subsection{Returns in the longshot and favorite groups}

Past tail demand predicts where wallets trade. We next ask whether it also
predicts the returns on those purchases. Table~\ref{tab:taildemandperformance} examines purchases made in the
month after classification in markets that subsequently resolved. We value
each purchase using the token's final payoff, as if the token had been held
until resolution. These terminal returns need not equal the wallet's actual
trading profit if the token was sold earlier.

\begin{table}[H]
	\centering
	\caption{Past tail demand and next-month terminal performance}
	\label{tab:taildemandperformance}
	\small
	\resizebox{0.98\linewidth}{!}{%
		\begin{tabular}{@{}lrrrr@{}}
			\toprule
			Purchases
			& \shortstack{Top-decile share of\\tail purchase dollars}
			& \shortstack{Terminal return:\\top decile}
			& \shortstack{Terminal return:\\other wallets}
			& \shortstack{Top-decile share of\\gross loss or net gain} \\
			\midrule
			Longshots & 26.6\% & $-21.96\%$ & $-20.33\%$ & 22.6\% of gross loss \\
			Favorites & 15.1\% & $+0.25\%$ & $+0.88\%$ & 4.8\% of net gain \\
			\bottomrule
	\end{tabular}}
	
	\medskip
	\begin{minipage}{0.95\linewidth}
		\footnotesize
		Notes: Groups are formed each month using purchases during the preceding six
		calendar months. Other wallets are the remaining 90 percent of the same
		monthly classification sample. Outcomes use purchases made in the following
		month in markets that subsequently resolved. Returns pool purchases within
		each group and compare the tokens' final payoffs with their purchase costs, as
		if the tokens had been held until resolution.
		
		Gross longshot losses are calculated separately within each group. For each
		wallet, we add the terminal gains and losses on its longshot purchases. A
		negative total enters gross losses in absolute value; a positive total
		contributes zero. For favorites, the final column reports the group's share of
		the aggregate net terminal gain on favorite purchases.
	\end{minipage}
\end{table}

The longshot group's pooled terminal return is $-21.96$ percent, close to the
$-20.33$ percent return for other wallets. It supplies 26.6 percent of
longshot purchase dollars but accounts for 22.6 percent of gross longshot
losses. Its share of losses is therefore not disproportionate to its share of
purchases. The favorite group earns 0.25 percent, compared with 0.88 percent
for other wallets. It supplies 15.1 percent of favorite purchase dollars but
receives only 4.8 percent of the aggregate net gain. Past demand predicts
where wallets trade much more clearly than who accounts for the returns at
the two price tails.

Table~\ref{tab:taildemandperformance} identifies the wallets that make the
purchases, but it does not establish whether they retain the tokens. We
therefore conduct a separate calculation. For each wallet's purchases of a
particular outcome token in one of the two tails, we check whether the same
wallet is later recorded selling that token before resolution. We retain
purchases for which no later sale is recorded. These tokens are more likely to
have remained with the buyer until resolution.\footnote{The restriction does
	not establish that the buyer held the tokens at resolution, because our data
	do not record every way in which a position can be transferred or changed.}

Panel A of Table~\ref{tab:no-later-sale} reports that the retained
longshot purchases cost \$99.7 million and were worth \$46.6 million
at resolution, producing an aggregate net loss of \$53.1 million. Wallets
classified in the longshot group when they first purchased that token below
10 cents account
for 10.5 percent of the amount paid but only 6.9 percent of the aggregate net
loss. Other wallets account for the remaining 93.1 percent. The corresponding favorite calculations are reported in Panel B.

\begin{table}[H]
	\centering
	\caption{Purchases with no later recorded sale by the buyer}
	\label{tab:no-later-sale}
	\small
	\resizebox{0.98\linewidth}{!}{%
		\begin{tabular}{@{}lrrrrr@{}}
			\toprule
			Buyer group
			& \shortstack{Purchase\\cost}
			& \shortstack{Value at\\resolution}
			& \shortstack{Gain or\\loss}
			& \shortstack{Share of\\purchase cost}
			& \shortstack{Share of aggregate\\net gain or loss} \\
			\midrule
			\multicolumn{6}{l}{\textit{Panel A. Longshots}} \\
			All wallets      & \$99.7M & \$46.6M & $-\$53.1$M & 100.0\% & 100.0\% \\
			Longshot group   & \$10.5M &  \$6.8M &  $-\$3.6$M &  10.5\% &   6.9\% \\
			Other wallets    & \$89.3M & \$39.8M & $-\$49.5$M &  89.5\% &  93.1\% \\
			\addlinespace
			\multicolumn{6}{l}{\textit{Panel B. Favorites}} \\
			All wallets      & \$5,778.5M & \$5,818.8M & $+\$40.3$M
			& 100.0\% & 100.0\% \\
			Favorite group   & \$834.4M & \$835.6M & $+\$1.2$M
			& 14.4\% & 2.9\% \\
			Other wallets    & \$4,944.1M & \$4,983.2M & $+\$39.1$M
			& 85.6\% & 97.1\% \\
			\bottomrule
	\end{tabular}}
	
	\medskip
	\begin{minipage}{0.95\linewidth}
		\footnotesize
		Notes: Longshots are purchases below 10 cents and favorites are purchases at
		or above 90 cents. Each panel retains purchases for which the buyer has no
		later recorded sale of the same outcome token before resolution. Value at
		resolution is the final value of the purchased tokens, and gain or loss is
		that value minus their purchase cost. Wallets are assigned to the longshot or
		favorite group according to their classification in the month of their first
		purchase of that token in the corresponding tail. The classification uses
		purchases made during the preceding six months. Other wallets include wallets
		outside the corresponding top decile and wallets that could not be classified
		in that month. The final column reports each group's gain or loss divided by
		the corresponding all-wallet total. It differs from the gross-loss measure in
		Table~\ref{tab:taildemandperformance}. The absence of a recorded sale does not
		establish that the buyer held the tokens until resolution because the data do
		not contain every way a position can be transferred or changed.
	\end{minipage}
\end{table}

Among purchases with no later recorded sale, recurrent longshot buyers account
for less of the aggregate net loss than their share of purchase cost.
The same pattern holds for favorites: the recurrent favorite group accounts
for 14.4 percent of purchase cost but only 2.9 percent of the aggregate net
gain. Both panels therefore show that recurrent tail buyers account for less
of the corresponding aggregate gain or loss than their share of purchase
cost.

\subsection{Are longshot losses concentrated among inexperienced wallets?}
\label{sec:experience}

Recurrent demand for a particular tail differs from general experience in
the market. We measure experience before each purchase month using purchase
dollars, active months, and the number of markets traded during the preceding
six months, together with the number of days since the wallet's first
observed trade. Within each month, we rank wallets on each of these four
measures, average the four ranks, and divide wallets into five equally sized
groups. This measures previous activity and time in the market, not an
independently observed level of skill.

All five experience groups lose on their longshot purchases. From the least-
to the most-experienced group, pooled returns are $-38.34$, $-44.13$,
$-23.68$, $-22.29$, and $-21.55$ percent. The ordering is not monotonic:
the second group loses more than the first, while returns become less negative
from the second through the fifth group. Parent-event-clustered 95 percent
confidence intervals exclude zero for every group's pooled return. Appendix
Table~\ref{tab:experience} reports the estimates and intervals.

More-experienced wallets may lose less simply because they trade in markets,
weeks, or price ranges where returns tend to be higher. To separate these
broad differences from the choices wallets make within them, we compare
returns across experience groups for purchases made in the same child market,
during the same calendar week, and at average prices within the same one-cent
range.\footnote{We retain only comparison groups containing purchases from at
	least two experience groups. Appendix~\ref{app:mechanisms} provides further
	details.} Within each comparison, wallets may still choose whether to buy YES
or NO, on which day to trade, and the exact price to pay. These choices can
affect returns. For each wallet, we measure the difference between its return
and the average return of all purchases in its comparison group. We then
average these differences within each experience group, weighting by purchase
dollars. The most-experienced group performs $0.83$ percentage point above its
comparison-group averages. This is a relative performance difference, not a
positive absolute return. Its 95 percent confidence interval is
$[0.22,1.44]$ percentage points. Among the other longshot groups, only group 3
has a within-comparison difference whose 95 percent interval excludes zero.

Longshot losses are not confined to inexperienced wallets. Among longshot
purchases for which the buyer has no later recorded sale of the same token,
the most-experienced group accounts for $58.2$ percent of gross losses. We
calculate these losses separately for each wallet and token: when the value
received at resolution is below the amount spent, we count the difference as
a loss, without using gains on other tokens to offset it.\footnote{
	Table~\ref{tab:taildemandperformance} instead combines gains and losses across
	all of a wallet's longshot tokens before determining whether that wallet made
	an overall loss. Table~\ref{tab:no-later-sale} reports the aggregate net loss,
	allowing gains on some purchases to offset losses on others.} The
most-experienced group accounts for most of the total dollars lost because it
buys substantially more longshots, not because it has the worst return per
dollar. In this same sample, its return is $-44.26$ percent, which is less
negative than the return of every other experience group.

Favorite purchases earn positive pooled returns in all five experience groups.
Returns are $+0.60$, $+0.66$, $+0.54$, $+0.65$, and $+0.84$ percent from the
least- to the most-experienced group. The first and fifth groups have 95
percent confidence intervals of $[0.18,1.03]$ and $[0.63,1.05]$ percent,
respectively. Their within-market-week-price differences are small: $-0.02$,
$-0.13$, $-0.03$, $-0.02$, and $+0.02$ percentage points. Among favorite
purchases with no later recorded sale, the most-experienced group accounts for
74.3 percent of purchase cost and 77.4 percent of aggregate net gains. Favorite
gains therefore are not confined to inexperienced wallets.

Taken together, the results distinguish persistence in trading choices from
the distribution of gains and losses. The recurrent longshot group loses
money, while the recurrent favorite group earns positive returns. Yet neither
group accounts for a disproportionate share of the corresponding losses or
gains: the recurrent longshot group's share of losses is below its share of
longshot purchases, and the recurrent favorite group earns less per dollar
than other wallets buying favorites. Past tail demand therefore predicts
where wallets subsequently trade more clearly than it predicts who ultimately
bears the losses or receives the gains. The experience results add that
longshot losses occur throughout the experience distribution and are not
mainly a novice phenomenon.

\FloatBarrier

\section{Interpreting the Tail Returns}
\label{sec:interpretation}

In the preceding sections, we document where tail returns occur and which
wallets repeatedly buy at the price tails. In this Section, we ask what these
findings imply for the two explanations that have organized the literature
on the \flb{}. The first explanation attributes the \flb{} to mistaken probability
assessments: bettors treat unlikely outcomes as more likely than they actually are and
likely outcomes as less likely, making longshots overpriced and favorites
underpriced \citep{griffith1949,tk1992,snowbergwolfers2010}. The second
explanation assumes that bettors assess the probabilities correctly but accept
lower expected returns because they value the small chance of a large payoff
\citep{weitzman1965,ali1977,julliensalanie2000}.
\citet{snowbergwolfers2010} pose the choice between these accounts as the
central question in this literature, and \citet{ottavianisorensen2008} survey
both alongside explanations based on heterogeneous beliefs and informed
trading.

The two accounts make claims about different objects. Misperception concerns
the relationship between prices and outcomes, and is therefore a statement
about probability errors rather than about returns. In turn, risk love and preference for skewness concern the traders who choose these payoffs, and are therefore
statements about who buys at the tails and what those buyers earn. Studies
that observe prices alone must infer preferences from pricing patterns.
Because we observe purchases together with the wallets that make them, we can
examine each claim on the appropriate object. We take the two accounts in turn and then consider the remaining explanations in light of Sections~\ref{sec:explanations} and~\ref{sec:traders}, asking which accounts are consistent with the
results.

\subsection{Misperception}
\label{sec:misperception}

Figure~\ref{fig:full-price-equal-market} shows probability errors across the
full range of purchase prices, giving each child market equal weight. Their
sign pattern is the one misperception predicts. From the bin beginning at
4 cents through the bin beginning at 45 cents, every probability error is
negative, meaning that tokens pay one dollar less often than their prices imply. From 50
cents through 99 cents, every probability error is positive. The confidence
intervals exclude zero in each of these bins.

The magnitudes, however, are largest away from the tails. The probability
error reaches $-5.86$ percentage points in the bin beginning at 35 cents and
$+7.28$ percentage points in the bin beginning at 60 cents. Within the
longshot tail it does not exceed $0.53$ percentage point in absolute value,
and at or above 95 cents it ranges from $+1.14$ to $+0.11$ percentage point.
Accounts based on the misperception of probabilities place their strongest
predictions at the smallest and largest probabilities, where the treatment of
unlikely outcomes departs most from the outcome frequencies. In these data the
departures are smallest there, and within the longshot tail they change sign between adjacent bins rather than describing a single direction.

The size of the longshot return also depends on the prices at which the
purchases occur. Longshot purchases cost \$273.6 million and bought 23.4
billion tokens, an average of 1.17 cents per token. At that average price the
pooled probability error of $-0.23$ percentage point corresponds to a return
of $-19.35$ percent. An account based on how bettors assess probabilities
therefore has to explain a difference of a few tenths of a percentage point
between prices and outcome frequencies, which becomes a large return only
after division by a low price.

Section~\ref{sec:generality} adds a further qualification. The two-sided
pattern is absent in Sports under both primary aggregations, and Sports
contains the largest number of listed markets in the sample. A general
feature of how bettors assess probabilities would be expected to appear
there.

Misperception therefore describes the relationship between prices and
outcomes across the interior of the price range, where the errors are large,
regular, and precisely estimated. It describes that relationship less well at
the two tails, which is where the \flb{} is conventionally measured.

\subsection{Risk love and preference for skewness}
\label{sec:risklove}

The second account implies that some traders repeatedly choose low-priced
claims and bear the losses that follow. Section~\ref{sec:traders} examines
both parts of that implication.

Recurrent demand exists. Wallets in the top decile of past longshot demand
supply 26.6 percent of the following month's purchases below 10 cents, and
they remain unusually likely to buy longshots after the classification
histories cease to overlap. Past purchases at low prices therefore identify
wallets that will buy at low prices again.

Their returns, however, resemble those of other buyers. The group's terminal return on
those purchases is $-21.96$ percent, against $-20.33$ percent for the
remaining wallets. Its share of gross longshot losses, 22.6 percent, is below
its 26.6 percent share of purchase dollars. Among purchases for which the
buyer has no later recorded sale, the group accounts for 10.5 percent of the
amount paid and 6.9 percent of the aggregate net loss. The corresponding
favorite group earns 0.25 percent against 0.88 percent for other wallets, and
receives 4.8 percent of the aggregate net gain from 15.1 percent of the
purchases.

Past demand therefore identifies where wallets trade more clearly than it
identifies which wallets bear the losses at those prices. The losses on
low-priced purchases are spread across buyers who do not repeatedly seek
them. An account in which the aggregate return pattern is sustained by an
identifiable group of traders with distinctive preferences receives no
support here, although the recurrent demand that such an account describes is
present in the data.

\subsection{Other explanations}
\label{sec:otheraccounts}

Neither of the two principal accounts describes the tail returns on its own.
Section~\ref{sec:explanations} limits explanations based on how purchases are
executed, on reported fees, and on shared collateral, and
Section~\ref{sec:experience} shows that the losses are not confined to
inexperienced wallets. Further explanations remain that these data do not
settle.

If the wallets selling low-priced tokens hold better information than those
buying them, buyers would lose to that information rather than to their own
beliefs or preferences \citep{ottavianisorensen2010}, and
\citet{gomezcram2026} document a small group of persistently informed traders
on Polymarket. We do not observe the information held by sellers. If traders
value holding a position on an event for its own sake, they would accept
lower returns without any error in assessing probabilities. We have no
measure of that motive.

The results are consistent with another explanation, in which prices at the
tails reflect the beliefs of the traders most willing to buy at those prices
rather than a distortion common to all bettors
\citep{ali1977,gandhiserrano2015}. Under that account the identity of the
marginal buyer changes from market to market, which would be consistent with
recurrent longshot buyers earning what other buyers earn and with losses
appearing at every level of experience. We do not test this account. Separating disagreement among buyers from the arrival of information and from differences in liquidity is not possible with transaction prices alone, and existing evidence on heterogeneous beliefs in betting markets rests on structural estimation \citep{gandhiserrano2015}.

\section{Conclusion}
\label{sec:conclusion}

Polymarket exhibits the classic favorite--longshot pattern in aggregate, but the pattern is not uniform across the platform. Purchases below 10 cents lose 6.3 cents per dollar when each separately listed contract receives equal weight, but earn 4.1 cents when related contracts are first grouped into broader events. Pooling all purchases produces an even larger longshot loss of 19.3 cents per dollar, as spending is concentrated in events with lower returns. The pattern also differs across categories: longshots lose and favorites gain in Crypto and Politics, whereas Sports does not show that pattern.

Past purchases strongly predict whether a wallet will continue buying at the same price tail, but is much less informative about which wallets gain or lose at those prices. Recurrent longshot buyers earn approximately the same returns as other longshot buyers and do not experience losses out of proportion to their spending; recurrent favorite buyers earn less than other favorite buyers. Longshot losses also persist after substantial previous trading, across experience levels and ways of buying, in markets without reported fees, and where collateral can be shared. These findings weigh against simple explanations based entirely on inexperienced buyers, one way of trading, reported fees, collateral costs, or a stable group of especially loss-prone longshot buyers. They do not, however, identify a unique explanation for the bias.

The broader lesson is that an aggregate FLB, how widely it appears across events, and which bettors carry its losses are separate empirical questions. In prediction markets that list many related contracts, aggregation is not merely a reporting choice: the return on dollars invested, the return for the average separately listed contract, and the return for the average broader event describe different features of the market. Persistent demand, likewise, need not reveal who ultimately receives the gains or bears the losses.

\clearpage
\bibliographystyle{plainnat}
\bibliography{references}

\clearpage
\appendix

\section{Tail Definitions}
\label{app:calibration}

\subsection{Symmetric tail thresholds}

\begin{figure}[H]
\centering
\includegraphics[width=0.92\linewidth]{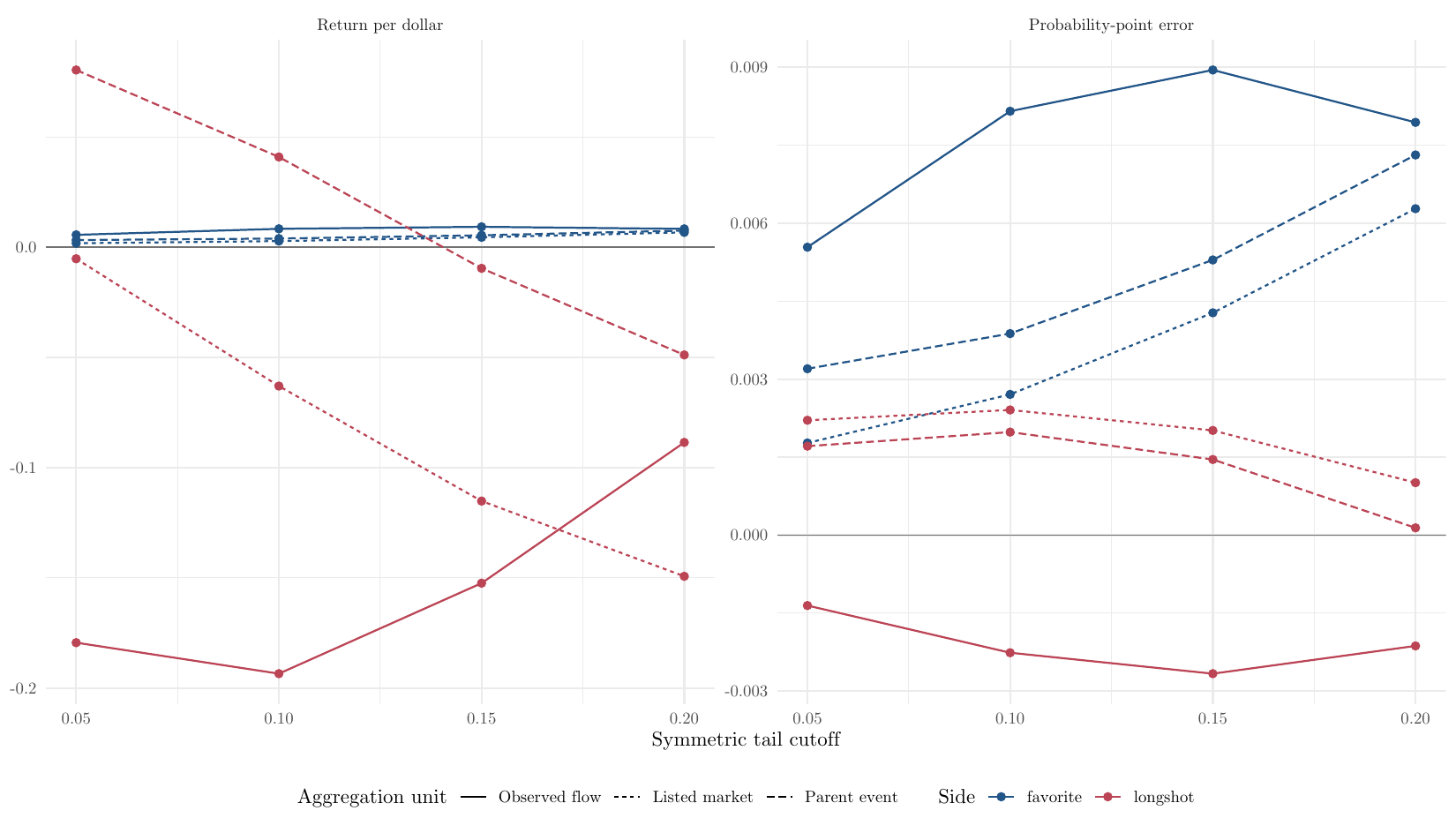}
\caption{Returns and probability errors across symmetric thresholds}
\label{fig:thresholds}
\begin{minipage}{0.94\linewidth}
\footnotesize
Notes: Longshots have $p<c$ and favorites $p\geq1-c$ for
$c\in\{0.05,0.10,0.15,0.20\}$. In the legend, ``Observed flow'' denotes
pooled purchases, ``Listed market'' gives each child market equal weight,
and ``Parent event'' gives each parent event equal weight. Within each
market or event, returns are weighted by purchase dollars and probability
errors by the number of tokens purchased, as in Section~\ref{sec:estimands}.
Both vertical scales use decimals.
\end{minipage}
\end{figure}

Figure~\ref{fig:thresholds} and Table~\ref{tab:thresholds} vary the longshot
cutoff from 5 to 20 cents and use the symmetric favorite cutoff. Pooled
longshot returns are negative and pooled favorite returns are positive at
every cutoff. The equal child-market longshot return is close to zero below
5 cents and becomes more negative as the cutoff rises. The equal parent-event
longshot return becomes negative at the 15/85 cutoff, but its confidence
interval excludes zero only at 20/80. Thus, the cutoff at which longshot
returns become negative depends on how purchases are grouped and weighted.

\begin{table}[H]
\centering
\caption{Symmetric threshold sensitivity}
\label{tab:thresholds}
\small
\begin{tabular}{@{}crrrr@{}}
\toprule
Cutoff & \shortstack{Pooled longshot\\return}
& \shortstack{Pooled favorite\\return}
& \shortstack{Equal child-market\\longshot return}
& \shortstack{Equal parent-event\\longshot return} \\
\midrule
5/95   & $-17.94\%$ & $+0.560\%$ & $-0.53\%$ & $+8.04\%$ \\
10/90  & $-19.35\%$ & $+0.833\%$ & $-6.30\%$ & $+4.09\%$ \\
15/85  & $-15.24\%$ & $+0.924\%$ & $-11.52\%$ & $-0.96\%$ \\
20/80  & $-8.86\%$  & $+0.831\%$ & $-14.93\%$ & $-4.89\%$ \\
\bottomrule
\end{tabular}

\medskip
\begin{minipage}{0.92\linewidth}
\footnotesize
Notes: Cutoffs are in cents: 5/95, for example, classifies purchases below
5 cents as longshots and purchases at or above 95 cents as favorites.
Pooled returns divide aggregate gains or losses at resolution by the amount
paid. The final two columns first calculate this ratio within each child
market or parent event, then average those returns equally. At the 5/95
cutoff, the 95 percent interval for the equal child-market longshot return
includes zero.
\end{minipage}
\end{table}

\subsection{Excluding endpoint prices}
\label{app:endpoint}

The additional restriction in Section~\ref{sec:exists} retains longshot
purchases above two cents and below ten cents, and favorite purchases at or
above 90 cents but below 99 cents. Table~\ref{tab:endpoint} reports the
results. This check removes purchases nearest the endpoints but does not
eliminate the general amplification of returns at low purchase prices.

\begin{table}[H]
\centering
\caption{Tail returns after excluding endpoint prices}
\label{tab:endpoint}
\small
\begin{tabular}{@{}llrr@{}}
\toprule
Tail & Aggregation & Return (\%) & 95\% interval\\
\midrule
Longshots & Equal child markets & $-4.12$ & $[-5.40,-2.84]$\\
Longshots & Pooled purchases & $-23.75$ & $[-34.69,-12.81]$\\
Favorites & Equal child markets & $0.187$ & $[0.109,0.265]$\\
Favorites & Pooled purchases & $1.729$ & $[1.250,2.209]$\\
\bottomrule
\end{tabular}\par
\medskip
\begin{minipage}{0.94\linewidth}
\footnotesize
Notes: Longshots satisfy $0.02<p<0.10$ and favorites satisfy
$0.90\leq p<0.99$. Intervals cluster by parent event. The table reports the
two primary aggregations.
\end{minipage}
\end{table}

\section{Time and Maturity}
\label{app:time}

\begin{figure}[H]
\centering
\includegraphics[width=0.94\linewidth]{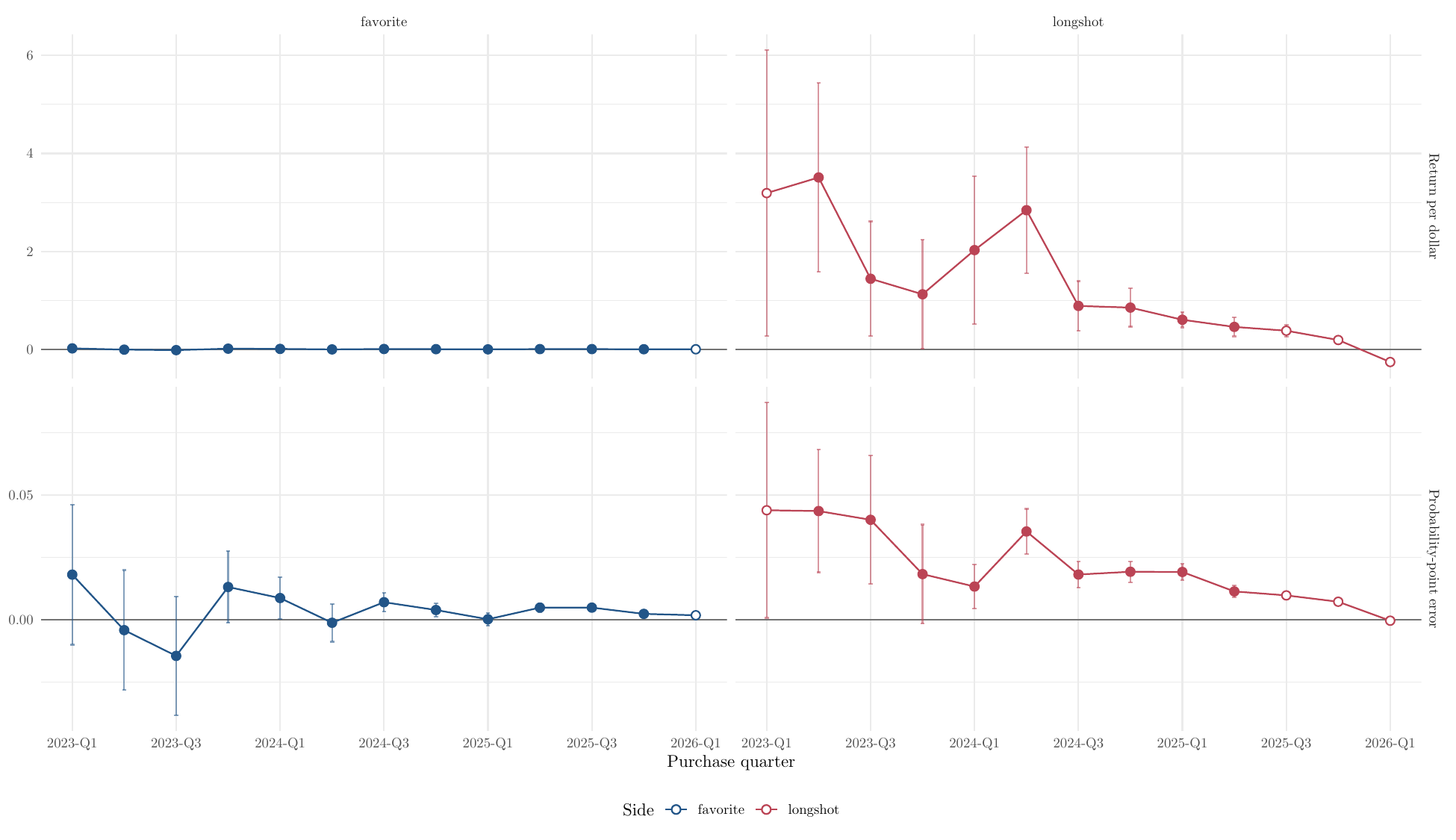}
\caption{Time variation and resolution coverage}
\label{fig:time}
\begin{minipage}{0.94\linewidth}
\footnotesize
Notes: Estimates give each child market equal weight within its purchase
quarter and tail; quarters with fewer than 50 represented child markets
are omitted. Both vertical scales use decimals. Intervals are 95 percent
confidence intervals clustered by parent event. Hollow markers indicate
that less than 95 percent of the amount paid in that quarter and tail has
a known final payoff by March 29, 2026. Those estimates remain provisional.
\end{minipage}
\end{figure}

Figure~\ref{fig:time} reports equal child-market returns and probability
errors by purchase quarter. Hollow markers identify quarters where less
than 95 percent of purchase dollars have a known final payoff by March 29,
2026. Their estimates cover only the purchases
that have resolved, so differences from earlier quarters may reflect which
markets resolve first. In particular, the negative returns among resolved
purchases in the first quarter of 2026 remain provisional.

Table~\ref{tab:maturity} restricts the sample by scheduled closing dates.
Pooled longshot returns remain negative in every row, while the signs of
the equal child-market and equal parent-event returns vary across these
different samples.

\begin{table}[H]
\centering
\caption{Longshot returns under restrictions on scheduled closing dates}
\label{tab:maturity}
\small
\begin{tabular}{@{}lrrr@{}}
\toprule
Restriction & \shortstack{Pooled\\return}
& \shortstack{Equal child-market\\return}
& \shortstack{Equal parent-event\\return} \\
\midrule
Full resolved sample & $-19.35\%$ & $-6.30\%$ & $+4.09\%$ \\
\shortstack[l]{Scheduled to close at least\\30 days before cutoff} & $-22.76\%$ & $+1.96\%$ & $+10.61\%$ \\
\shortstack[l]{Scheduled to close at least\\90 days before cutoff} & $-24.44\%$ & $+30.66\%$ & $+27.83\%$ \\
\shortstack[l]{Scheduled close no later than\\1 day after purchase} & $-10.86\%$ & $-13.14\%$ & $-0.90\%$ \\
\shortstack[l]{Scheduled close no later than\\7 days after purchase} & $-15.12\%$ & $-7.48\%$ & $+3.13\%$ \\
\shortstack[l]{Scheduled close no later than\\30 days after purchase} & $-29.90\%$ & $+2.39\%$ & $+12.11\%$ \\
\bottomrule
\end{tabular}

\medskip
\begin{minipage}{0.94\linewidth}
\footnotesize
Notes: All rows use resolved longshot purchases below 10 cents. The second and third rows retain markets scheduled to close at least 30 or 90 days before March 29, 2026. In the final three rows, purchases must have occurred at least 1, 7, or 30 days before that cutoff, respectively, and the market must have been scheduled to close no later than the corresponding number of days after purchase. Returns pool purchases, average child-market returns equally, or average parent-event returns equally, as indicated.
\end{minipage}
\end{table}

\section{Event Structure}
\label{app:eventstructure}

\subsection{Parent-event-weighted returns}
\label{app:parent-weighting}

Let $\mathcal{G}$ denote the set of parent events. For each parent event
$g\in\mathcal{G}$, let $\mathcal{M}_g\subseteq\mathcal{M}$ contain the child
markets listed under that event heading. Not every parent event is represented
in both tails. Let:
\[
\mathcal{G}_T
=
\left\{
g\in\mathcal{G}:
\mathcal{M}_g\cap\mathcal{M}_T\neq\varnothing
\right\}
\]
denote the parent events with at least one child market in $\mathcal{M}_T$.
The condition
$\mathcal{M}_g\cap\mathcal{M}_T\neq\varnothing$ therefore means that parent
event $g$ contains at least one child market with purchases in tail $T$.

For each parent event in $\mathcal{G}_T$, we calculate a return using the
tail-$T$ purchases from all its child markets:
\begin{equation}
	R_{gT}
	=
	\frac{
		\sum_{m\in\mathcal{M}_g\cap\mathcal{M}_T}
		\sum_{f\in\mathcal{F}_{mT}}q_f(y_f-p_f)
	}{
		\sum_{m\in\mathcal{M}_g\cap\mathcal{M}_T}
		\sum_{f\in\mathcal{F}_{mT}}p_fq_f
	},
	\qquad
	g\in\mathcal{G}_T,\quad T\in\{L,H\}.
	\label{eq:parent-roi}
\end{equation}
The numerator adds the gains and losses across the child markets listed under
parent event $g$, and the denominator adds the amounts paid in those markets.
Their ratio, $R_{gT}$, is the return on the event's purchases in tail $T$,
taken together. Within a parent event, child markets with more spending
contribute more to $R_{gT}$.

We then average these returns across the parent events represented in the
tail:
\begin{equation}
	R_T^P
	=
	\frac{1}{|\mathcal{G}_T|}
	\sum_{g\in\mathcal{G}_T}R_{gT},
	\qquad
	T\in\{L,H\}.
	\label{eq:equal-parent}
\end{equation}
Here, $|\mathcal{G}_T|$ is the number of represented parent events. The sum
contains one return for each event, so every parent event receives the same
weight in $R_T^P$. A parent event containing many child markets receives the
same weight as an event containing only one, and the amount spent in the event
does not affect its weight in the final average.

\subsection{Aggregation decomposition}
\label{app:aggregation-decomposition}

For the longshot tail, let $G$ be the number of represented parent events,
$n_g$ the number of represented child markets in parent $g$, and
$N=\sum_g n_g$. Let $d_g$ be the dollars paid for longshot purchases in that
parent and $D=\sum_g d_g$. Finally, let $e_g$ be the equal-child-market return
within parent $g$ and $p_g$ its pooled-purchase return. The four calculations
used in Section~\ref{sec:aggregation-change} can then be written as
\begin{align*}
R_{\mathrm{child}}&=\sum_g\frac{n_g}{N}e_g,
&
R_{\mathrm{bridge}}&=\sum_g\frac{1}{G}e_g,\\
R_{\mathrm{parent}}&=\sum_g\frac{1}{G}p_g,
&
R_{\mathrm{pooled}}&=\sum_g\frac{d_g}{D}p_g.
\end{align*}
The difference between the equal-child-market and equal-parent-event returns
is exactly
\begin{align*}
R_{\mathrm{parent}}-R_{\mathrm{child}}
&=\left(R_{\mathrm{bridge}}-R_{\mathrm{child}}\right)
  +\left(R_{\mathrm{parent}}-R_{\mathrm{bridge}}\right)\\
&=\sum_g\left(\frac{1}{G}-\frac{n_g}{N}\right)e_g
  +\sum_g\frac{1}{G}(p_g-e_g)\\
&=10.329+0.063=10.392\ \text{percentage points}.
\end{align*}
The first term changes only the weights across parent events; the second
changes only how purchases are combined within each parent event. The
difference between the platform-wide pooled return and the equal-parent-event
return is
\begin{align*}
R_{\mathrm{pooled}}-R_{\mathrm{parent}}
&=\sum_g\left(\frac{d_g}{D}-\frac{1}{G}\right)p_g\\
&=-23.440\ \text{percentage points}.
\end{align*}
This identity changes only the weights across parent events, from equal
weights to weights based on dollars paid.

\subsection{Returns by number of child markets}

A child market is one separately traded question within a parent event.
For example, a parent event about an election can contain a different child
market for each candidate. Table~\ref{tab:menusize} groups parent events by
this count, not by the number of tokens traded or the number of purchases.
It supplements the aggregation decomposition in
Section~\ref{sec:aggregation-change}; it is not an estimate of the effect of
adding another child market. Returns vary nonmonotonically across the groups.

\begin{table}[H]
\centering
\caption{Longshot returns by number of child markets in the parent event}
\label{tab:menusize}
\small
\resizebox{0.98\linewidth}{!}{%
\begin{tabular}{@{}lrrrrr@{}}
\toprule
\shortstack{Child markets\\per parent} & \shortstack{Parent\\events} & \shortstack{Represented\\child markets} & \shortstack{Share of\\amount paid} & \shortstack{Equal child\\market return} & \shortstack{Equal parent\\event return} \\
\midrule
1      & 198,231 & 198,231 & 23.1\% & $-2.34\%$  & $-2.34\%$  \\
2--5   & 23,247  & 79,140  & 15.3\% & $+39.51\%$ & $+47.26\%$ \\
6--10  & 15,020  & 116,429 & 11.9\% & $-16.05\%$ & $+8.93\%$  \\
11--25 & 8,106   & 96,123  & 27.3\% & $-15.40\%$ & $+32.32\%$ \\
26+    & 1,836   & 86,161  & 22.4\% & $-34.16\%$ & $-11.64\%$ \\
\bottomrule
\end{tabular}}

\medskip
\begin{minipage}{0.94\linewidth}
\footnotesize
Notes: Parent events are grouped by their total number of listed child markets.
The represented-child-market column counts child markets with resolved
longshot purchases in each group. Share of amount paid is the group's share
of total spending on resolved longshot purchases. The child-market count
used to group events includes listed child markets without resolved
longshot purchases. The final two columns give equal weight to represented
child markets or parent events within each row, respectively.
\end{minipage}
\end{table}

\section{Trading Conditions}
\label{app:tradingconditions}

This appendix gives the sample definitions, return calculations, and
confidence intervals for the trading-condition comparisons in
Section~\ref{sec:explanations}.

\subsection{Previous trading, buyer action, and reported fees}

Previous trading dollars are accumulated from earlier purchase months in the
same child market, using purchases across all price ranges in the resolved
analysis sample. The current month is excluded. Observations are child
market by month by price bin. Within category, month, and tail, these observations
are ordered by previous trading dollars and divided into four groups. Thus,
a child market can enter more than one price-bin observation; the grouping
is not a division of distinct markets into four permanent categories. The
return reported in Panel A of Table~\ref{tab:tradingconditions} pools
purchases within each group. Confidence intervals are not available for
these four groups.

A buyer posting a purchase offer is recorded as the maker; a buyer accepting
a seller's existing offer is recorded as the taker. Table~\ref{tab:executionappendix}
reports both tails and both primary aggregations. A market carries a fee
flag when its metadata indicate a fee or a positive base fee. The calculations
do not subtract transaction-specific fees or include rebates.

\begin{table}[H]
\centering
\caption{Tail returns by buyer action and reported fee: confidence intervals}
\label{tab:executionappendix}
\small
\setlength{\tabcolsep}{3.2pt}
\begin{tabular}{@{}llrrrr@{}}
\toprule
& & \multicolumn{2}{c}{Equal child markets} & \multicolumn{2}{c}{Pooled purchases}\\
\cmidrule(lr){3-4}\cmidrule(l){5-6}
Group & Tail & Return & 95\% interval & Return & 95\% interval\\
\midrule
\multicolumn{6}{l}{\textit{Panel A. Buyer action}}\\
Posted purchase offer & Longshots & $-3.50$ & $[-5.69,-1.31]$ & $-17.56$ & $[-29.44,-5.69]$\\
 & Favorites & $0.641$ & $[0.607,0.675]$ & $0.816$ & $[0.608,1.025]$\\
Accepted seller's offer & Longshots & $-31.17$ & $[-32.94,-29.40]$ & $-22.17$ & $[-86.22,41.88]$\\
 & Favorites & $-0.460$ & $[-0.515,-0.405]$ & $0.877$ & $[0.544,1.210]$\\
\addlinespace
\multicolumn{6}{l}{\textit{Panel B. Reported fee}}\\
No fee flag & Longshots & $-5.20$ & $[-7.54,-2.86]$ & $-20.78$ & $[-49.71,8.16]$\\
 & Favorites & $0.260$ & $[0.221,0.298]$ & $0.864$ & $[0.618,1.111]$\\
Fee flag & Longshots & $-13.03$ & $[-16.88,-9.17]$ & $10.03$ & $[-4.30,24.35]$\\
 & Favorites & $0.371$ & $[0.283,0.458]$ & $0.333$ & $[-0.047,0.712]$\\
\addlinespace
\bottomrule
\end{tabular}\par
\medskip
\begin{minipage}{0.96\linewidth}
\footnotesize
Notes: All numbers are percentages. Returns compare terminal payoffs with
amounts paid before fees. Confidence intervals cluster by parent event.
The reported fee flag identifies markets whose metadata indicate a fee;
it does not identify the fee actually paid on each transaction. These
intervals describe each group separately and are not tests of differences
between groups.
\end{minipage}
\end{table}

\subsection{Transaction prices after posted purchase offers}
\label{app:postpurchase}

The construction first combines purchases made through posted offers for each
outcome token and five-minute purchase period. The average purchase price is
total dollars paid divided by tokens purchased. The resulting purchase group
is classified as a longshot or favorite using that average price.

For each group, target times are five minutes, one hour, and one day after
the start of the purchase period. The target must precede resolution. The
calculation selects the first later five-minute price interval whose timestamp
is no earlier than the target and no later than 10 minutes, one hour, or six
hours after it, respectively. It uses that interval's closing transaction
price, requiring a price strictly between zero and one. It therefore does
not use a synchronized quote or the first individual trade after the target.
Horizons are measured from the start of the purchase period, not from the exact
time of every purchase within it.

For each horizon, the price change is multiplied by the number of tokens
purchased and summed over eligible purchase groups; dividing by their purchase
cost gives the reported return. Terminal returns instead use all resolved
purchases made through posted offers, classified by each purchase's price.
The terminal observations therefore need not match the interim groups,
even apart from availability of subsequent prices.

\begin{table}[H]
\centering
\caption{Available purchase samples at each return horizon}
\label{tab:postpurchase}
\small
\setlength{\tabcolsep}{5pt}
\begin{tabular}{@{}llrrr@{}}
\toprule
Tail & Horizon & \shortstack{Five-minute\\purchase groups} & Amount paid & Return (\%)\\
\midrule
Longshots & 5 minutes & 9,970,723 & \$127.01M & $+3.92$\\
& 1 hour & 11,571,901 & \$117.47M & $+16.91$\\
& 1 day & 10,761,733 & \$88.34M & $+7.56$\\
& Resolution & --- & \$167.65M & $-17.56$\\
\addlinespace
Favorites & 5 minutes & 3,192,378 & \$5.174B & $+0.169$\\
& 1 hour & 3,624,304 & \$4.958B & $+0.252$\\
& 1 day & 2,493,150 & \$2.308B & $+0.263$\\
& Resolution & --- & \$7.000B & $+0.816$\\
\bottomrule
\end{tabular}\par
\medskip
\begin{minipage}{0.95\linewidth}
\footnotesize
Notes: A five-minute purchase group combines purchases of the same token
through posted offers within that period. Amount paid is the total cost of
the purchases included at the stated horizon. The resolution rows classify
individual purchases by price and therefore have no directly comparable
group count. The number of groups need not decrease at each successive
horizon because the windows used to find later prices differ.
No interim confidence intervals are available for these horizon-specific
samples. Terminal intervals appear in Table~\ref{tab:executionappendix};
Table~\ref{tab:postpurchasebalanced} reports intervals for the common sample.
The sample and timing rules above apply to Figure~\ref{fig:postpurchaseprices}.
\end{minipage}
\end{table}

\begin{table}[H]
\centering
\caption{Posted-offer returns in the common four-horizon sample}
\label{tab:postpurchasebalanced}
\small
\setlength{\tabcolsep}{5pt}
\begin{tabular}{@{}llrr@{}}
\toprule
Tail & Horizon & Return (\%) & 95\% confidence interval\\
\midrule
Longshots & 5 minutes & $+4.85$ & $[4.08,5.62]$\\
& 1 hour & $+6.96$ & $[5.46,8.46]$\\
& 1 day & $+6.62$ & $[2.14,11.10]$\\
& Resolution & $-36.27$ & $[-60.88,-11.66]$\\
\addlinespace
Favorites & 5 minutes & $+0.017$ & $[-0.003,0.037]$\\
& 1 hour & $+0.014$ & $[-0.020,0.047]$\\
& 1 day & $+0.308$ & $[0.187,0.429]$\\
& Resolution & $+1.939$ & $[1.378,2.499]$\\
\bottomrule
\end{tabular}

\medskip
\scriptsize
\begin{tabular}{@{}lrrrrr@{}}
\toprule
Tail & \shortstack{Five-minute\\purchase groups} & Purchases & Child markets & Parent events & Amount paid\\
\midrule
Longshots & \shortstack{6,526,943\\(41.3\%)}
& \shortstack{20,088,147\\(30.5\%)}
& \shortstack{50,258\\(9.6\%)}
& \shortstack{13,815\\(6.9\%)}
& \shortstack{\$64.62M\\(37.7\%)}\\
Favorites & \shortstack{1,179,457\\(20.7\%)}
& \shortstack{4,138,636\\(14.2\%)}
& \shortstack{25,403\\(6.3\%)}
& \shortstack{9,825\\(4.6\%)}
& \shortstack{\$1.710B\\(25.6\%)}\\
\bottomrule
\end{tabular}\par
\medskip
\begin{minipage}{0.96\linewidth}
\footnotesize
Notes: The common sample retains five-minute purchase groups formed from
posted offers with an eligible
price at all three interim horizons and a terminal payoff. Tail status is set
once from the group's average purchase price and then held fixed. The
confidence intervals cluster by parent event. Parentheses in the coverage
panel divide each count or dollar amount by its corresponding total across
all reconstructed purchase groups in the same tail. Purchases count the
individual purchases within those groups. Among
groups whose one-day target precedes resolution, the common sample represents
67.7 percent of longshot dollars and 66.8 percent of favorite dollars.
\end{minipage}
\end{table}

Table~\ref{tab:postpurchase} uses a different available sample at each
horizon, so its rows do not trace the returns of a fixed set of purchases.
Table~\ref{tab:postpurchasebalanced} follows the same purchase groups at
every horizon. Both comparisons nevertheless use transaction prices, which
may depend on whether the later trade was initiated by a buyer or seller.
An increase in those prices need not be an increase in the price at which
the original buyer could sell. Moreover, purchases with later recorded
trading may differ from those without it. The common sample resolves the
changing-sample problem but does not make these comparisons a definitive
test of outdated purchase offers.

We do not split the last transaction in each five-minute period according
to whether the buyer or seller initiated it. Such a split would still use
transaction prices and would not establish the price at which the original
buyer could sell.

\subsection{Shared collateral and metadata coverage}
\label{app:collateral}

Shared collateral applies to events with mutually exclusive outcomes that
jointly cover the event's possible resolution. A NO token for one outcome
can be converted into YES tokens for the other outcomes. In the three-candidate
example, a dollar first creates A-YES and A-NO; converting A-NO gives B-YES
and C-YES, leaving a complete set backed by the original dollar. This
description abstracts from any conversion fee. The conversion rule is
documented in
\href{https://github.com/Polymarket/neg-risk-ctf-adapter}{Polymarket's published contract documentation}.

The verified metadata audit covers 590,449 of 591,187 market identifiers in
the resolved purchase sample, a classification rate of 99.88 percent. The
remaining 738 identifiers are unclassified rather than assigned to the
no-shared-collateral group. Table~\ref{tab:collateralcoverage} reports both
tails under the same verified-metadata restriction. It reproduces the returns
in Panel D of Table~\ref{tab:tradingconditions} and adds market and event
counts, amounts paid, and confidence intervals for each row.

\begin{table}[H]
\centering
\caption{Shared collateral under verified metadata coverage}
\label{tab:collateralcoverage}
\scriptsize
\setlength{\tabcolsep}{3pt}
\resizebox{0.99\linewidth}{!}{%
\begin{tabular}{@{}llrrrll@{}}
\toprule
Tail & Shared collateral & Child markets & Parent events & Amount paid
& Equal child-market return (95\% CI) & Pooled return (95\% CI)\\
\midrule
Longshots & No  & 464,826 & 227,563 & \$114.22M
& $-5.94\;[-8.15,-3.72]$ & $-9.79\;[-22.27,2.69]$\\
Longshots & Yes & 110,550 & 18,635 & \$158.38M
& $-7.80\;[-13.36,-2.24]$ & $-25.76\;[-72.36,20.84]$\\
Favorites & No  & 352,502 & 211,176 & \$6.363B
& $+0.182\;[0.142,0.222]$ & $+0.588\;[0.366,0.809]$\\
Favorites & Yes & 89,689 & 17,487 & \$3.246B
& $+0.644\;[0.568,0.721]$ & $+1.286\;[0.813,1.759]$\\
\bottomrule
\end{tabular}%
}\par
\medskip
\begin{minipage}{0.96\linewidth}
\footnotesize
Notes: Returns and confidence intervals are percentages. ``No'' denotes a
verified absence of the shared-collateral flag; missing metadata are excluded,
not coded as no shared collateral. Equal child-market returns give each represented
child market equal weight. Pooled returns divide aggregate terminal gains or
losses by aggregate purchase cost. Confidence intervals cluster by parent
event and describe each cell separately rather than testing differences
between collateral regimes.
\end{minipage}
\end{table}

Shared collateral need not affect only trading against high prices. It may
also change which events are listed or attract different traders. The
cross-sectional comparison is consequently not a causal test of collateral
costs. The available transactions also do not record all conversions,
creations, combinations, redemptions, and off-exchange transfers of tokens.
They cannot reconstruct all collateral committed or the complete positions
of traders exploiting price differences across outcomes. Prices observed
at different times cannot establish that all components of such a trade
were simultaneously available.

\subsection{Conditional comparisons}

Table~\ref{tab:architecturecontrols} reports descriptive regressions for both
tails. Each observation combines purchases in one child market, purchase
month, and price bin. Regressions give these observations equal weight and
include category-by-quarter fixed effects, price-bin indicators, and duration
categories.

The shared-collateral coefficient is interacted with Crypto, Politics, and
Sports. Its base coefficient therefore applies only to the other categories.
For a displayed interaction category, the corresponding association is the
sum of the base and additional coefficients. The base coefficient is
therefore not a platform-wide difference. Standard errors for the sums are
not reported.

\begin{table}[H]
\centering
\caption{Conditional associations with tail probability errors and returns}
\label{tab:architecturecontrols}
\small
\setlength{\tabcolsep}{4pt}
\begin{tabular}{@{}lrrrr@{}}
\toprule
& \multicolumn{2}{c}{Longshots} & \multicolumn{2}{c}{Favorites}\\
\cmidrule(lr){2-3}\cmidrule(l){4-5}
Variable & \shortstack{Probability\\error} & Return & \shortstack{Probability\\error} & Return\\
\midrule
Log (1 + child-market count) & \shortstack{$-0.134$\\($0.030$)} & \shortstack{$-11.184$\\($1.428$)} & \shortstack{$-0.112$\\($0.039$)} & \shortstack{$-0.117$\\($0.041$)}\\
\addlinespace[.4em]
Log (1 + previous purchase dollars) & \shortstack{$-0.170$\\($0.012$)} & \shortstack{$-8.727$\\($0.694$)} & \shortstack{$0.084$\\($0.012$)} & \shortstack{$0.088$\\($0.012$)}\\
\addlinespace[.4em]
Shared collateral: base coefficient & \shortstack{$-0.909$\\($0.221$)} & \shortstack{$-34.821$\\($10.595$)} & \shortstack{$0.544$\\($0.229$)} & \shortstack{$0.577$\\($0.241$)}\\
\addlinespace[.4em]
Additional coefficient: Crypto & \shortstack{$1.130$\\($0.258$)} & \shortstack{$39.691$\\($12.877$)} & \shortstack{$-0.586$\\($0.267$)} & \shortstack{$-0.622$\\($0.281$)}\\
\addlinespace[.4em]
Additional coefficient: Politics & \shortstack{$-0.017$\\($0.314$)} & \shortstack{$-0.342$\\($15.284$)} & \shortstack{$-0.295$\\($0.319$)} & \shortstack{$-0.315$\\($0.335$)}\\
\addlinespace[.4em]
Additional coefficient: Sports & \shortstack{$0.270$\\($0.239$)} & \shortstack{$19.741$\\($11.730$)} & \shortstack{$0.281$\\($0.252$)} & \shortstack{$0.291$\\($0.265$)}\\
\addlinespace[.4em]
\bottomrule
\end{tabular}\par
\medskip
\begin{minipage}{0.96\linewidth}
\footnotesize
Notes: Estimates and parent-event-clustered standard errors (in parentheses)
are multiplied by 100; both outcome columns are in percentage-point units.
Probability errors compare payoffs with purchase prices; returns divide
gains or losses by purchase cost. Logarithms use one plus the stated
variable. Child-market count is the number listed under the parent event;
previous purchase dollars are accumulated in the child market before the
current month. The base shared-collateral coefficient applies outside Crypto,
Politics, and Sports; the three additional coefficients modify it within
those categories. The longshot sample contains 2,326,485 market-month-price
observations and 246,198 parent events; the favorite sample contains
1,588,365 observations and 228,663 parent events. All displayed variables
enter jointly with duration categories, price-bin indicators, and
category-by-quarter fixed effects. These are associations, not causal effects.
\end{minipage}
\end{table}

\section{Classifying Wallets by Past Tail Demand}
\label{app:traders}

\subsection{Construction of the classifications}

For wallet $i$ and measurement month $m$, define the share of purchase dollars
spent on longshots during the preceding six calendar months as
\begin{equation}
  s^L_{i,m-1}=\frac{\sum_{t=m-6}^{m-1}\sum_{f\in(i,t)}
  p_fq_f\mathbf{1}\{p_f<0.10\}}
  {\sum_{t=m-6}^{m-1}\sum_{f\in(i,t)}p_fq_f}.
  \label{eq:propensityappendix}
\end{equation}
For each month, the classification includes wallets with positive purchases
over the preceding six calendar months, activity in at least two of those
months, and the sample screens in Section~\ref{sec:data}. We regress the
longshot share on log prior purchase dollars, active months, log market
count, and Politics, Sports, and Crypto purchase shares. We rank wallets by
the difference between their actual longshot share and the share predicted by
these characteristics. The highest decile forms the longshot group. We
construct the favorite group in the same way after replacing the numerator in
Equation~\eqref{eq:propensityappendix} with purchases at or above 90 cents.
The classifications therefore identify wallets that direct more of their
purchase dollars to the relevant tail than their scale, breadth, trading history, and
category mix would predict. Neither classification uses future purchases or
returns.

\begin{table}[H]
\centering
\caption{Alternative prospective longshot classifications}
\label{tab:prospectiverobust}
\small
\setlength{\tabcolsep}{4pt}
\begin{tabular*}{\textwidth}{@{\extracolsep{\fill}}lrrrr@{}}
\toprule
Classification & \shortstack{Distinct\\wallets}
& \shortstack{Share of next-month\\longshot spending}
& \shortstack{Longshot share of the\\group's next-month\\spending}
& \shortstack{Longshot minus\\middle-price return\\(percentage points)} \\
\midrule
Rolling six months & 102,399 & 26.65\% & 1.544\% & $-23.66$ \\
Expanding window & 88,760 & 26.40\% & 1.785\% & $-25.99$ \\
\shortstack[l]{Odd/even-month\\cross-fit} & 115,867 & 18.10\% & 2.767\% & $-31.43$ \\
\bottomrule
\end{tabular*}

\medskip
\begin{minipage}{0.94\linewidth}
\footnotesize
Notes: Rows identify the top decile of past longshot demand. The expanding
window uses all prior months; odd/even cross-fitting uses prior odd months
for even measurement months and vice versa. Distinct wallets count addresses
that enter the group at least once. The first spending share divides the
group's longshot spending by all classified wallets' longshot spending in
the next month. The second divides the group's longshot spending by its
spending across all prices. The final column subtracts the group's pooled
middle-price return from its pooled longshot return, in percentage points,
before controls for markets and periods. Table~\ref{tab:taildemandperformance}
instead compares longshot returns across wallet groups.
\end{minipage}
\end{table}

\begin{figure}[H]
\centering
\includegraphics[width=0.94\linewidth]{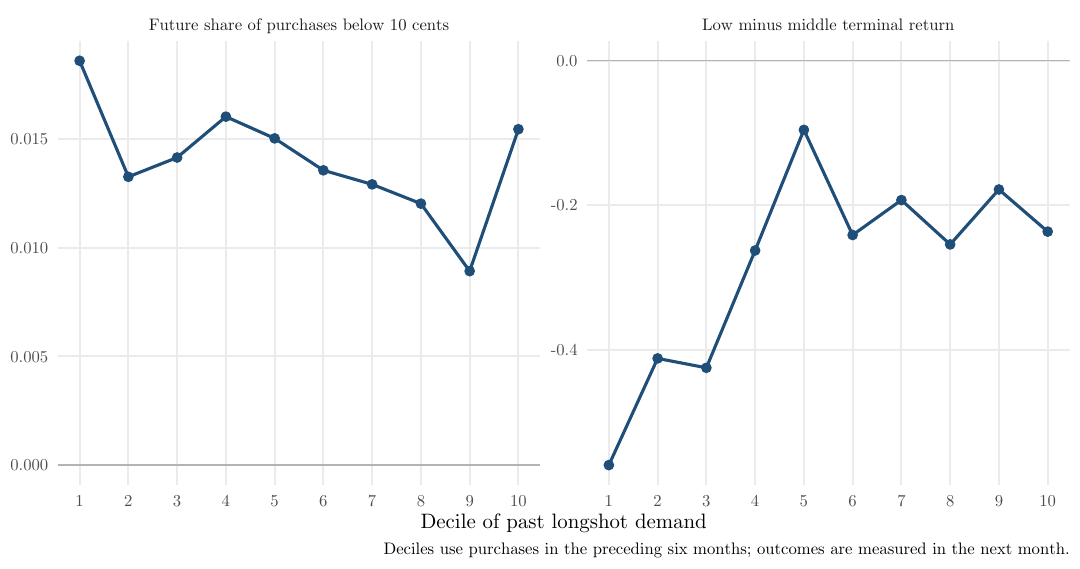}
\caption{Past longshot demand, future demand, and future performance}
\label{fig:doseappendix}
\begin{minipage}{0.92\linewidth}
\footnotesize
Notes: Deciles rank the difference between each wallet's actual six-month
longshot spending share and the share predicted by its prior trading activity
and category mix. The left panel reports the next-month share of each decile's
purchase dollars spent below 10 cents; the right reports its pooled longshot
return minus its pooled middle-price return. Both vertical scales use decimals.
\end{minipage}
\end{figure}

In Figure~\ref{fig:doseappendix}, neither the share of next-month spending
on longshots nor the longshot-minus-middle-price return varies monotonically
across deciles. The classification ranks past longshot demand relative to
trading activity and category mix; it need not rank the unadjusted future
spending shares shown here.

\subsection{Trader profile}

Table~\ref{tab:profileappendix} compares wallets classified in March 2026
using the dataset's full-sample wallet characteristics. The median of wallets'
average time from purchase to resolution is 110.4 days in the longshot group,
compared with 19.3 days among other classified wallets. Longshot-group wallets
trade fewer markets, have larger median trades, trade more heavily in Sports
and less in Crypto, and have a similar maker share. These differences describe
where and how the groups trade; they do not establish why their returns differ.

\begin{table}[H]
\centering
\caption{Median wallet characteristics, March 2026}
\label{tab:profileappendix}
\small
\begin{tabular}{@{}lrr@{}}
\toprule
Characteristic & Longshot group & Other classified wallets \\
\midrule
Average purchase-to-resolution time (days) & 110.4 & 19.3 \\
Markets traded & 26 & 47 \\
Median trade size (dollars) & 26.06 & 20.94 \\
Sports share & 22.2\% & 7.0\% \\
Crypto share & 0.4\% & 7.9\% \\
Maker share & 13.2\% & 12.0\% \\
\bottomrule
\end{tabular}

\medskip
\begin{minipage}{0.92\linewidth}
\footnotesize
Notes: Group membership is assigned in March 2026; characteristics come from
the dataset's full-sample wallet summaries. Entries are medians across wallets.
The first row is the median of wallets' average purchase-to-resolution times;
the trade-size row is the median of wallets' median trade sizes. Sports,
Crypto, and maker shares use the wallet-level measures supplied with the
dataset. Makers post offers that other traders accept. Other classified
wallets are the remaining nine deciles of past longshot demand.
\end{minipage}
\end{table}

\subsection{Overlap between the longshot and favorite groups}

The one-month correlations in the favorite and longshot rankings are 0.855
and 0.849, respectively. Adjacent-month top-decile retention is 64.5 and 68.5
percent, although the six-month measurement windows overlap. The longshot and
favorite rankings have a Spearman correlation of $-0.313$. Only 45
wallet-month observations belong to both top deciles, out of 412,565 that
belong to at least one.

\begin{figure}[H]
\centering
\includegraphics[width=0.78\linewidth]{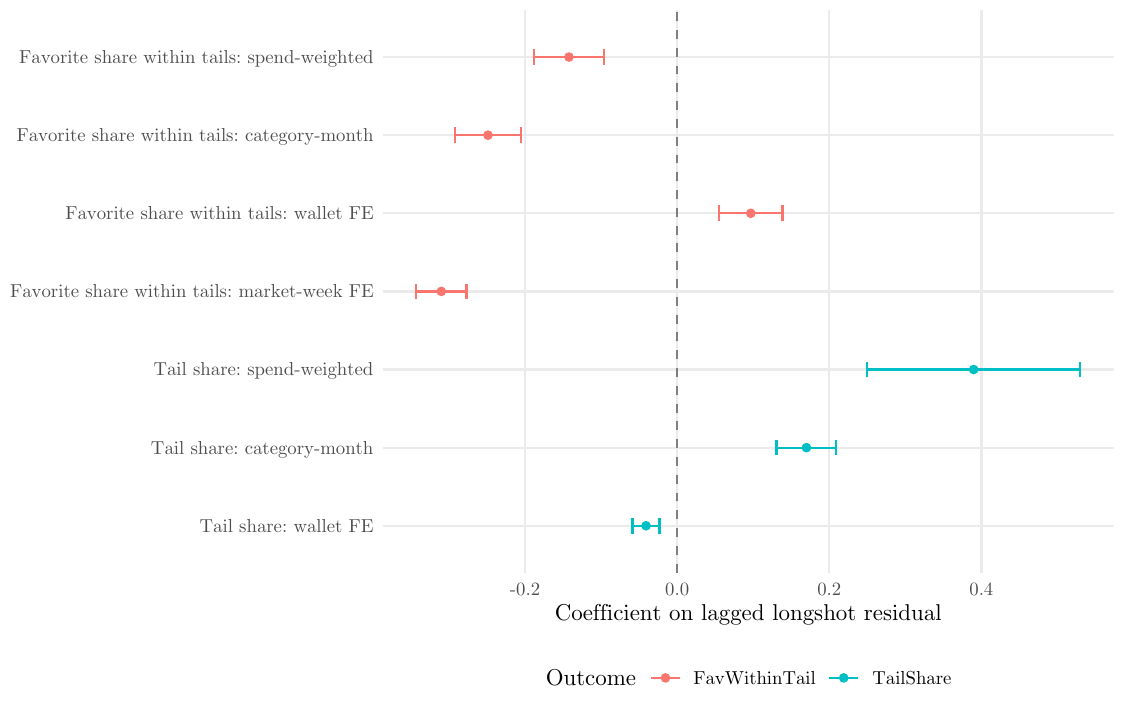}
\caption{Past longshot demand and future purchases}
\label{fig:pastdemandcoefficients}
\begin{minipage}{0.92\linewidth}
\footnotesize
Notes: The explanatory variable is the prior difference between a wallet's
actual longshot spending share and the share predicted by its trading
activity and category mix. The outcomes are the next-month share of all
purchase dollars spent on longshots or favorites, and the favorite share
of those tail dollars. Longshots are priced below 10 cents and favorites at
or above 90 cents. Points are regression coefficients and horizontal bars
are 95 percent confidence intervals. The baseline is the category-month
specification. ``FE'' denotes fixed effects: wallet fixed effects compare
observations for the same wallet over time. Spending-weighted specifications
weight observations by dollars spent in the tails.
\end{minipage}
\end{figure}

Figure~\ref{fig:pastdemandcoefficients} distinguishes the share of all
purchase dollars spent in either tail from the share of tail dollars spent
on favorites. In the baseline comparison across wallets, higher past
longshot demand predicts a larger share spent in the tails and a smaller
favorite share within them. The associations reverse in the specification
with wallet fixed effects, which compares changes within the same wallet.
The relationship across wallets therefore does not describe how an
individual wallet's purchases change over time.

Table~\ref{tab:favoritemirror} uses a common sample for spending shares,
net-gain shares, and favorite returns: purchases with a known final payoff,
a recorded closing or expiration date, and a price at least 90 cents but
strictly below one dollar. The favorite group's spending share is 15.5
percent in this sample. The 15.1 percent demand share reported in the main
text uses the broader purchase sample, including unresolved purchases and
purchases at one dollar.

\begin{table}[H]
\centering
\caption{Past favorite demand and return capture}
\label{tab:favoritemirror}
\small
\begin{tabular*}{\textwidth}{@{\extracolsep{\fill}}lrrr@{}}
\toprule
Group or ranking
& \shortstack{Share of favorite\\purchase dollars}
& \shortstack{Share of aggregate net\\gains on favorites}
& \shortstack{Favorite\\return} \\
\midrule
Favorite group & 15.5\% & 4.8\% & 0.25\% \\
\addlinespace
Neither group & 64.5\% & 76.4\% & 0.97\% \\
Longshot-only group & 20.0\% & 18.8\% & 0.77\% \\
Favorite-only group & 13.9\% & 4.2\% & 0.25\% \\
Both groups & 1.6\% & 0.6\% & 0.32\% \\
\addlinespace
\shortstack[l]{Past-favorite-demand\\decile 1} & n/a & n/a & 1.39\% \\
\shortstack[l]{Past-favorite-demand\\decile 10} & n/a & n/a & 0.25\% \\
\bottomrule
\end{tabular*}

\medskip
\begin{minipage}{0.94\linewidth}
\footnotesize
Notes: All columns use the sample described above. Spending shares divide
the group's favorite purchase cost by the sample total of \$6.385 billion;
net-gain shares divide its aggregate gain at resolution by the sample total
of \$52.188 million. Returns divide each group's net gain by its purchase
cost. The first row includes wallets in the favorite top decile, whether or
not they are also in the longshot top decile. The next four rows partition
wallet-months by membership in those two groups and sum to 100 percent in
each share column, apart from rounding. The final two rows compare the lowest
and highest deciles of past favorite demand. Classification uses purchases
before the measurement month.
\end{minipage}
\end{table}

\section{Additional Wallet Return Calculations}
\label{app:incidence}

\subsection{Broader realized trading outcomes}

The broader wallet calculations include trades outside the longshot purchases
used for the main return comparison. An observed purchase can be matched to
a later recorded sale of the same outcome token, matching the earliest
unmatched purchases first. These completed buy--sell pairs earn
$+\$11.222$ million on $\$121.459$ million of matched purchase cost.
Separately, wallets ever classified in the longshot group earn
$+\$117.958$ million in the dataset's resolved whole-wallet profit measure.
Neither calculation is a return on a pure buy-and-hold longshot strategy.
They show why the terminal loss attributed to a longshot purchase need not
equal its buyer's overall trading loss.

\begin{table}[H]
\centering
\caption{Broader realized trading outcomes}
\label{tab:broaderpnlappendix}
\small
\begin{tabular}{@{}lrrr@{}}
\toprule
Calculation & Purchase base or sample & Profit & Return\\
\midrule
Completed buy--sell pairs & \$121.459M matched cost & $+\$11.222$M & 9.24\%\\
Whole-wallet outcomes & 68,145 ever-classified wallets & $+\$117.958$M & n/a\\
\bottomrule
\end{tabular}\par
\medskip
\begin{minipage}{0.94\linewidth}
\footnotesize
Notes: The first row matches observed purchases to later same-token sales,
using the earliest unmatched purchases first. Its purchase base is the cost
of the quantities matched to those sales, and its return divides matched
sale proceeds minus that cost by the cost. The second row uses the dataset's
resolved whole-wallet profit for wallets ever classified in the longshot
group, including their trading outside longshots. Neither row measures the
return to holding only longshots until resolution. Wallets are addresses,
not necessarily distinct people.
\end{minipage}
\end{table}

\subsection{Purchases with and without a later recorded sale}

Table~\ref{tab:incidenceappendix} separates resolved longshot purchases
according to whether their buyer is later recorded selling the same token
before resolution. It does not match individual purchased tokens to later
sales or reconstruct final ownership.

Purchases assigned to buyers with a later same-token sale have an aggregate
terminal gain of \$0.215 million, close to zero. Purchases with no later
recorded sale have an aggregate terminal loss of \$53.145 million. Wallets
outside the longshot group when they first bought that token below ten cents
account for 93.1 percent of the latter loss.

\begin{table}[H]
\centering
\caption{Longshot purchases and later sales by their buyers}
\label{tab:incidenceappendix}
\small
\setlength{\tabcolsep}{4pt}
\begin{tabular}{@{}lrrrr@{}}
\toprule
Purchase group & Amount paid & Terminal gain/loss & Return & Share\\
\midrule
All resolved longshots & \$273.577M & $-\$52.930$M & $-19.35\%$ & 100.0\%\\
Buyer later sells same token & \$173.864M & $+\$0.215$M & $+0.12\%$ & 63.6\% of dollars\\
No later recorded sale & \$99.713M & $-\$53.145$M & $-53.30\%$ & 36.4\% of dollars\\
\quad Longshot group at first purchase & \$10.453M & $-\$3.643$M & $-34.85\%$ & 6.9\% of net loss\\
\quad Other wallets at first purchase & \$89.260M & $-\$49.502$M & $-55.46\%$ & 93.1\% of net loss\\
\bottomrule
\end{tabular}\par
\medskip
\begin{minipage}{0.96\linewidth}
\footnotesize
Notes: Terminal gain or loss is the tokens' payoff at resolution minus
purchase cost, irrespective of subsequent sale. The indented rows partition
purchases with no later recorded sale using the buyer's classification in the
month of its first longshot purchase of that token. Other wallets include
wallets outside the top decile and unclassified wallets. The loss shares
refer to aggregate net loss, not gross losses. In the final column, the first
three rows report shares of total resolved longshot spending; the indented
rows report shares of the aggregate net loss on purchases with no later
recorded sale. The absence of a recorded
sale does not establish final ownership: the data omit some token
creations, combinations, redemptions, and transfers.
\end{minipage}
\end{table}

For favorites with no later recorded sale, Table~\ref{tab:no-later-sale}
reports \$5.779 billion of purchase cost and a \$40.3 million aggregate net
gain. The recurrent favorite group accounts for 14.4 percent of that cost but
2.9 percent of the gain. As in the longshot panel, these spending shares
do not establish final ownership because the data omit some ways positions can
be transferred or changed.

\section{Wallet Experience}
\label{app:mechanisms}

\subsection{Classification before the purchase month}

For each purchase month, experience is based on four measures known before
that month: purchase dollars during the preceding six calendar months,
the number of active months in that period, the number of distinct child
markets traded during it, and days since the wallet's first observed trade.
The experience classification includes wallets with positive purchase dollars
over the preceding six months. Unlike the tail-demand classification, it
does not require activity in at least two of those months.

Wallets are ranked separately on each measure within the month. The four
percentile ranks receive equal weight, and their average determines five
equally sized groups. Group 1 has the lowest average rank and group 5 the
highest. Ties are broken by wallet address. The
classification changes monthly and does not use the current month's returns.

\subsection{Absolute returns and comparisons within markets}

Table~\ref{tab:experience} reports longshots in Panel A and favorites in
Panel B. The first return column pools purchases in the measurement month.
The comparison-difference column subtracts the pooled return of purchases
in the same child market, calendar week, and one-cent average-price range.
Only comparison groups represented by at least two experience groups are
used. The remaining return differences are averaged using purchase dollars
as weights.

For example, a wallet return of $-18$ percent against a comparison-group
return of $-20$ percent produces a difference of $+2$ percentage points.
A positive number in that column therefore does not mean a positive absolute
return. The comparison removes between-group differences in markets, weeks,
and average-price bins, but it does not isolate a causal effect of experience.
Exact prices and purchase composition can still differ within a bin.

\begin{table}[H]
\centering
\caption{Wallet experience and tail-purchase returns}
\label{tab:experience}
\scriptsize
\setlength{\tabcolsep}{3pt}
\resizebox{0.99\linewidth}{!}{%
\begin{tabular}{@{}clllcc@{}}
\toprule
\shortstack{Experience\\group}
& \shortstack{Pooled return\\(\%; 95\% CI)}
& \shortstack{Comparison difference\\(pp; 95\% CI)}
& \shortstack{No-later-sale return\\(\%; 95\% CI)}
& \shortstack{No-later-sale\\spending share (\%)}
& \shortstack{No-later-sale\\outcome share (\%)}\\
\midrule
\multicolumn{6}{l}{\textit{Panel A. Longshots; final column is share of gross losses}}\\
1 & $-38.34\;[-57.32,-19.36]$ & $-1.05\;[-3.33,1.23]$
  & $-64.02\;[-76.47,-51.57]$ & 4.76 & 4.81\\
2 & $-44.13\;[-62.85,-25.41]$ & $-1.11\;[-2.46,0.25]$
  & $-68.12\;[-76.48,-59.77]$ & 7.76 & 7.83\\
3 & $-23.68\;[-35.21,-12.16]$ & $-2.66\;[-4.74,-0.57]$
  & $-51.50\;[-59.89,-43.10]$ & 13.09 & 13.11\\
4 & $-22.29\;[-36.97,-7.61]$ & $-0.68\;[-1.52,0.16]$
  & $-49.63\;[-55.11,-44.16]$ & 16.04 & 16.07\\
5 & $-21.55\;[-33.80,-9.30]$ & $+0.83\;[0.22,1.44]$
  & $-44.26\;[-48.84,-39.69]$ & 58.36 & 58.19\\
\addlinespace
\multicolumn{6}{l}{\textit{Panel B. Favorites; final column is share of net gains}}\\
1 & $+0.60\;[0.18,1.03]$ & $-0.024\;[-0.082,0.034]$
  & $+0.14\;[-0.29,0.56]$ & 2.19 & 0.45\\
2 & $+0.66\;[0.12,1.20]$ & $-0.129\;[-0.256,-0.002]$
  & $+0.46\;[0.18,0.74]$ & 4.08 & 2.83\\
3 & $+0.54\;[0.22,0.86]$ & $-0.032\;[-0.066,0.003]$
  & $+0.69\;[0.47,0.91]$ & 8.11 & 8.39\\
4 & $+0.65\;[0.39,0.91]$ & $-0.021\;[-0.057,0.016]$
  & $+0.64\;[0.51,0.77]$ & 11.34 & 10.95\\
5 & $+0.84\;[0.63,1.05]$ & $+0.016\;[0.005,0.026]$
  & $+0.70\;[0.61,0.78]$ & 74.28 & 77.39\\
\bottomrule
\end{tabular}%
}\par
\medskip
\begin{minipage}{0.96\linewidth}
\footnotesize
Notes: Group 1 is least experienced and group 5 most experienced by the
prospective rank described above. The difference column uses only
market-week-price comparison groups with at least two experience groups and
is measured in percentage points, not absolute returns. The final three
columns use a separate sample of purchases with no later recorded same-token
sale before resolution. Spending shares divide each group's amount paid by
the total for that tail in this separate sample. Longshot outcome shares use gross losses, so positive
wallet-token outcomes do not offset negative ones; favorite outcome shares use
aggregate net gains. All confidence intervals cluster by parent event.
\end{minipage}
\end{table}

\subsection{Purchases with no later recorded sale}

The final three columns use purchases of a token by a wallet with no subsequent
recorded sale of that token before resolution. Experience is assigned in the
month when the wallet first purchases that token in the corresponding tail.
Terminal gains and losses are added for each wallet and token separately.
Negative totals contribute their absolute value to gross losses; positive
totals contribute zero to the longshot gross-loss measure. Gross losses are
then summed within longshot experience groups; the favorite panel instead
reports each group's share of aggregate net gains. This differs from the
wallet-level gross loss in
Table~\ref{tab:taildemandperformance}, where results are netted across the
wallet's longshot tokens, and from the aggregate net loss in
Table~\ref{tab:no-later-sale}, where gains offset losses.

The highest-experience group has a less negative return per dollar but
accounts for the largest share of gross losses. The group trades more
purchase dollars, so a larger loss share need not indicate worse
performance per dollar. Neither the raw return sequence nor the within-market
difference sequence is monotonic. The absence of a later sale does not
establish that the wallet retained the token until resolution.
For favorites, the highest-experience group likewise dominates purchase
dollars: it accounts for 74.3 percent of cost and 77.4 percent of net gains in
the classified no-later-sale sample. Its return is $+0.70$ percent, compared
with $+0.14$ percent for the least-experienced group; the latter interval
includes zero. Thus the concentration of dollars in experienced wallets is
present in both tails, while the sign of their terminal returns differs.

\section{Data Quality and Reconciliation}
\label{app:dataquality}

The released data contain 2,480,104 wallet characteristic rows and the same
number of profit-and-loss rows. Five profit measures---base, resolved,
no-fee, spread-adjusted, and spread-adjusted resolved---sum to zero across
the full wallet population to
absolute deviations below $10^{-5}$ dollars. An independent reconstruction of
resolved signed terminal gains and losses for a deterministic sample of 9,781 wallets
has a Spearman correlation of 0.9926 with the released wallet summary, agrees
in sign for every wallet in the top one percent by absolute reconstructed profit or loss,
and reconciles within one dollar for 96.4 percent of the sample.

\begin{table}[H]
\centering
\caption{Volume reconciliation}
\label{tab:volumeappendix}
\small
\begin{tabular}{@{}lr@{}}
\toprule
Accounting step & Dollar amount \\
\midrule
Wallet-level volume, both sides & $\$134.28$B \\
One-sided transaction equivalent & $\$67.14$B \\
One-sided equivalent after wallet wash filter & $\$64.03$B \\
Purchases after transaction and market restrictions & $\$24.62$B \\
Purchases after buyer counterparty-HHI restriction & $\$23.67$B \\
Purchases in resolved markets & $\$22.49$B \\
Longshot purchases in resolved markets & $\$0.274$B \\
\bottomrule
\end{tabular}

\medskip
\begin{minipage}{0.90\linewidth}
\footnotesize
Notes: The first row adds the dataset's wallet-level trading volume, which
counts both sides of a transfer. The second divides it by two, and the third
applies the wallet wash-trading filter before dividing by two. The final four
rows report amounts paid by buyers in the constructed purchase sample, with
the successive restrictions described in Table~\ref{tab:sample-construction}. The final
row restricts resolved purchases to prices below 10 cents.
\end{minipage}
\end{table}

Official Gamma metadata cover 590,449 of 591,187 market identifiers in the
resolved purchase sample. Events with verified shared collateral account for
58.1 percent of covered longshot purchase dollars. Appendix
Section~\ref{app:collateral} documents this comparison and the remaining
favorite calculations. Table~\ref{tab:architecturecontrols} reports
conditional associations, including the category interactions needed to
interpret the shared-collateral coefficient. Excluding the largest,
five largest, or ten largest parent events changes pooled longshot return
from $-19.35$ percent to $-15.8$, $-14.5$, and $-10.6$ percent,
respectively. The full-sample equal parent-event return is $+4.09$ percent. This
sensitivity motivates Webb-bootstrap inference for the pooled returns. The
null-imposed $p$-values are 0.184 for longshots and 0.001 for favorites.

\end{document}